\documentclass[fleqn,usenatbib]{mnras}

\usepackage[T1]{fontenc}

\DeclareRobustCommand{\VAN}[3]{#2}
\let\VANthebibliography\thebibliography
\def\thebibliography{\DeclareRobustCommand{\VAN}[3]{##3}\VANthebibliography}

\usepackage{graphicx}	
\usepackage{amsmath}	
\usepackage{amssymb}	
\usepackage{multirow}
\usepackage{newtxtext,newtxmath}

\title[Pulsar Kicks and Galactic Gamma-rays]{
Millisecond Pulsar Kicks Cause Difficulties in Explaining the Galactic Center Gamma-Ray Excess}

\author[O. Boodram $\&$ C. Heinke]{
Oliver Boodram,$^{1,2}$\thanks{E-mail: olbo7449@colorado.edu}
Craig O. Heinke$^{1}$
\\
$^{1}$Department of Physics, CCIS 4-183, University of Alberta,  Edmonton, AB, T6G 2E1, Canada\\
$^{2}$ Department of Aerospace Engineering Sciences, University of Colorado Boulder, 3775 Discovery Dr, Boulder, CO, 80303, USA\\
}

\begin{document}
\label{firstpage}
\pagerange{\pageref{firstpage}--\pageref{lastpage}}
\maketitle

\begin{abstract}
The unexplained excess gamma-ray emission from the Milky Way's Galactic Center has puzzled astronomers for nearly a decade. Two theories strive to explain the origin of this excess: self-annihilating dark matter particles or an unresolved population of radio millisecond pulsars. We examine the plausibility of a pulsar origin for the GeV excess using N-body simulations. We 
simulated millisecond pulsars in a realistic dynamical environment: (i) 
pulsars were born from the known stellar mass components of our Galaxy; (ii) 
pulsars were given natal velocity kicks 
as empirically observed from two different studies (or, for comparison, without kicks);
(iii)  
pulsars 
were evolved in a Galactic gravitational potential consistent with observations. Multiple populations of 
pulsars  
(with different velocity kicks)
were simulated over 1 Gyr. 
With final spatial distributions of pulsars
, we constructed synthetic gamma-ray surface brightness profiles. 
From comparisons with 
published Fermi-LAT surface brightness profiles, 
our pulsar simulations 
cannot reproduce the concentrated emission in the central 
degrees of the Bulge, though models without natal velocity kicks approach the data. 
We considered additive combinations of our (primordial MSP) simulations with models where pulsars are deposited from destroyed globular clusters in the Bulge, and a simple model for pulsars produced in the nuclear star cluster. 
We can reasonably reproduce the measured central gamma-ray surface brightness distribution of Horiuchi and collaborators using several combinations of these models, but we cannot reproduce the measured  distribution of Di Mauro with any combination of models. 
Our fits provide constraints on potential pathways to explain the gamma-ray excess using MSPs. 
\end{abstract}




\section{Introduction}

Following observations of the Milky Way 
bulge 
with the Fermi Gamma-ray Space Telescope (Fermi-LAT), many analyses revealed an unexpected excess of GeV gamma-ray emission 
\citep{Goodenough09,Vitale09,Abazajian12,Ajello16,Ackermann17}.
This excess could not be accounted for by previously modeled astrophysical backgrounds, such as cosmic ray interactions with molecular clouds 
\citep[e.g.][]{Macias14}.
Although alternatives have been suggested \citep{Carlson14,Cholis15b}, most explanations of the Galactic Center Excess (GCE) invoke dark matter annihilation or an unresolved population of millisecond radio pulsars (MSPs).

 Annihilation of dark matter particles in regions of high density has been suggested to create a gamma-ray signature similar to that observed 
\citep{Goodenough09,Hooper11,Hooper11b}.
More specifically, the spatial morphology of the GCE can be well described with the annihilation of weakly interacting massive particles (WIMPs) following a Navarro Frenk-White (NFW) density profile, as expected for the dark matter distribution in the Milky Way's bulge 
\citep[e.g.][]{Abazajian12,diMauro21}.
The GCE 
spectrum appears consistent with 
some models for 
WIMP annihilation into Standard Model particles
\citep{Hooper11b,Cerdeno15}.

Alternatively, 
an unresolved population of millisecond radio pulsars (MSPs), rapidly rotating neutron stars, 
in the GC could produce this excess emission
\citep{Abazajian11,Abazajian12,Yuan14}. 
MSPs are produced by accretion in low-mass X-ray binaries (LMXBs). In a stellar binary, after one of the stellar companions undergoes a supernova and produces a neutron star, and assuming the stellar binary remains intact after such a disruptive event, the neutron star left behind accretes material from its low mass companion star forming an LMXB \citep{Bhattacharya91}. The transfer of angular momentum in the process "spins up" the neutron star to millisecond rotation periods, and an MSP is left behind \citep{Alpar82,Archibald09,Papitto13}. 
Now, some studies have shown 
similarities between the spatial profile of LMXBs in M31 (which should be similar to their descendants, MSPs) and an NFW density profile, such that either might explain the GCE morphology 
\citep{Yuan14,Eckner18}.
Furthermore, the typical gamma-ray spectrum 
of MSPs, 
as measured from individual MSPs \citep{Abdo09} or from globular clusters \citep{Abdo10}, where MSPs are highly overabundant  \citep{Camilo05},
can resemble that of the GCE \citep{Abazajian11,Abazajian12}.
The number of MSPs required to explain the GCE has been estimated at 10,000-20,000 \citep{Yuan14}, 2,000-14,000 \citep{Cholis15}, $\sim$40,000 \citep{Ploeg17}, or $\sim$11,000 \citep{Gonthier18}. 
Substantial recent discussion has been spent on whether the shape of the GCE is better described by the Galactic bulge stellar distribution, or by a more spherical distribution  \citep{Macias18,Bartels18,Macias19,diMauro21},
and/or whether it shows evidence for the "bumpiness" expected from a stochastic distribution of MSP gamma-ray luminosities   
\citep{Lee16,Leane19,
Buschmann20,Leane20}.

An important question is the origin of these MSPs, which could be produced through normal binary evolution by the stars of the Galactic bulge, or through dynamical interactions in globular clusters and/or the nuclear stellar cluster. MSPs are believed to be of order 100 times more common in dense globular clusters than in the Galaxy as a whole, as their progenitors the LMXBs are, due to dynamical interactions \citep{Phinney94,Hui11,Bahramian13}. Since of order half of all the globular clusters initially present in our Galaxy are thought to have spiraled in to destruction in the inner galaxy \citep{Gnedin14}, \citet{Brandt15} suggested that these destroyed globular clusters formed the MSPs to produce the GCE, explored in more detail by e.g. \citet{Abbate18,Eckner18,Fragione18}.
\citet{Yuan14} suggested that MSPs are formed dynamically through interactions on large scales in the Bulge, following the discovery by \citet{Voss07a} that there is an excess of LMXBs per unit mass in the central 1' ($\sim$230 pc) of M31. 
\citet{Faucher-Giguere11} estimate that of order 1000 MSPs might be produced in the nuclear star cluster through dynamical interactions, scaling from the properties and pulsar content of the dense globular cluster Terzan 5. 
\citet{Macias19} prefers (on the basis of preferred matching of the GCE with the boxy bulge morphology) a "normal" (primordial binary) origin of the MSPs producing the GCE.  Most prior works did not analyze the effects of supernova kicks on the positions of primordial MSPs, though \citet{Eckner18} used a simple smoothing function to roughly approximate the effect. As we were completing this draft, we became aware of \citet{Ploeg21} which performs a somewhat similar analysis, with a different emphasis.

In this work, we 
attempt to constrain the kinematics of pulsars that might explain the GCE radial distribution. 
Specifically, we aim to 
explore the effects on the GCE of 
natal velocity  kicks received by newly formed neutron stars due to asymmetries in their supernovae explosions. 
We compare N-body simulations of MSPs with different 
velocity kick prescriptions, 
compared to 
the 
GCE radial distribution as seen 
by Fermi-LAT. 

To assess this issue, 
we simulate 
the dynamical evolution of a population of NSs retained in binaries in the Galactic gravitational potential. (These binaries will eventually come into contact as LMXBs, and then later appear as MSPs; as the binaries reach their new orbits on timescales of $\sim10^8$ years, compared to the $10^9$ and $10^{10}$ year lifetimes of LMXBs and MSPs respectively, we assume LMXBs and MSPs as having equivalent dynamical properties.)
We initiate the binaries subject to our understanding of the stellar mass 
distribution of 
our Galaxy, using prescriptions for the initial velocity kick of the binaries (due to the formation of the NS) motivated from empirical observations of LMXBs and MSPs. 
We evolve the binaries under the influence of a realistic Milky Way galactic potential. 
Using the galactic dynamics Python package, \textit{galpy}
\citep{Bovy15},
and its Runge Kutta integrator in C, equations of motion for each LMXB/MSP are integrated forward in time, implementing the Milky Way potential into Hamilton's equations, and the trajectories of the MSPs are solved for 1 Gyr of evolution. From the final positions of these MSPs, we are able to compare the MSP spatial distribution 
to the spatial morphology of the GCE detected by FermiLAT.  
This analysis has the potential to constrain or exclude an origin of the GCE produced by MSPs formed according to the stellar mass of our Galaxy (a primordial MSP origin), though we also explore contributions from MSPs produced by destroyed globular clusters and the nuclear star cluster (dynamical MSP origins). 
Section 2 demonstrates our treatment of initial conditions for a Galactic population of LMXBs/MSPs. Section 3 shows our interpretation of our simulated data. Finally, Section 4 holds the conclusion and summary of our work.

\section{Methods and Simulations}
We used 
Galactocentric cylindrical coordinates where positions are measured with respect to the center of the Milky Way Galaxy for our numerical simulations. Hence, the phase space elements of each MSP in our simulation are described by the time evolution of six quantities: $r, \phi, z, v_r, v_{\phi}, v_z$. Throughout our work, we assumed a circular velocity of 220 km/s at the solar radius of 8.2 kpc 
\citep{Bland-Hawthorn16}.
These solar parameters appropriately scale the units in \textit{galpy} to physical units such as kpc.

As stellar binaries form in the various mass components of our Galaxy, some binaries contain a low-mass and high-mass main sequence star. Eventually, the high mass star will undergo a supernova. The infalling material onto the proto-neutron star is not spherically symmetric which leads to asymmetries in the shock producing the supernova. As such, the proto-neutron star is spit out from the SNR with a rather significant, randomly oriented velocity comparable to, if not larger than, the star's intrinsic orbital velocity. Assuming the binary still remains intact after such a disruptive event, the stellar pair will have been supplied with additional velocity from the supernova known as a natal kick velocity. Thus, it is 
thought that LMXBs (and therefore their descendants MSPs) should receive kicks, altering their velocities from their birth velocities, and indeed these are seen \citep{Brandt95,Jonker04,Repetto17}. 

We sample natal kick velocities for each MSP from a Maxwellian distribution (\ref{eq:maxwellian})
\citep{Hobbs05}:
\begin{equation}
    P(v)=v^2e^{\frac{-v^2}{2\sigma^2}}
	\label{eq:maxwellian}
\end{equation}
where this sampling probability relies on $v$, a given velocity component, and $\sigma$, the 1D rms. We applied three natal kick prescriptions to our population of LMXBs/MSPs. 
We applied two kick prescriptions based on empirical observations of MSP velocities \citep{Hobbs05}, and of neutron star LMXB velocities \citep{Repetto17}, respectively (since LMXBs evolve into MSPs, we expect equivalence).
Both studies employ a Maxwellian distribution for natal kicks, where the latter advocates for higher natal kicks than the former. 
 While it is not impossible that MSPs in the Bulge might have a different kick distribution than elsewhere in the Galaxy, we are not aware of any rationale pointing toward such a scenario.
For reference, we also examined a population which did not receive any natal kicks whatsoever.

\begin{figure}
	\includegraphics[width=\columnwidth]{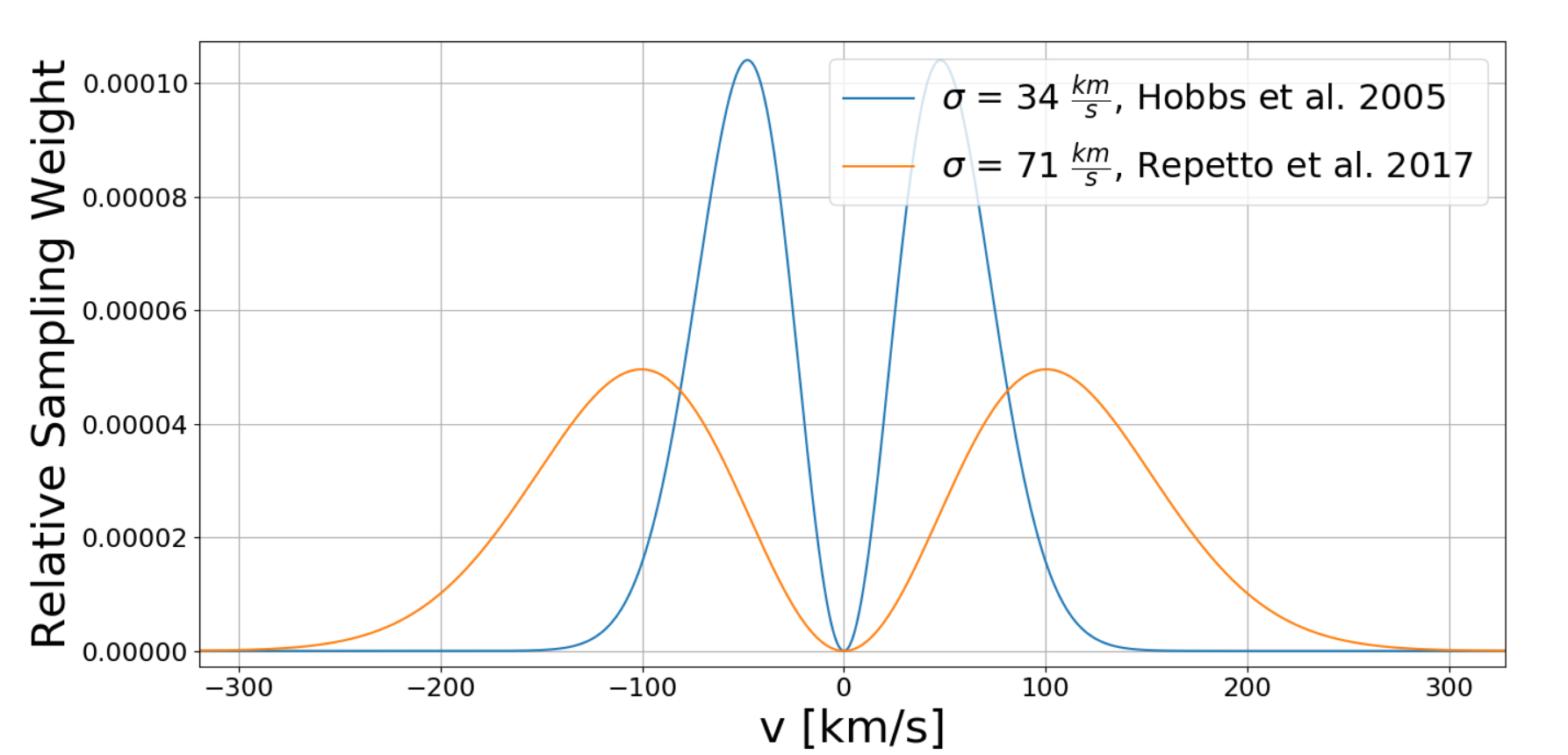}
    \caption{
    1D Maxwellian velocity distributions for natal kick prescriptions. 
    \citet{Hobbs05} prefers a lower kick description of LMXBs, whereas 
    \citet{Repetto17} 
    prefers higher kick velocities.}
    \label{fig:maxwellian_kicks}
\end{figure}

Sampling each component from (\ref{eq:maxwellian}), a $\vec{v}_{kick} = (v_{r, kick}, v_{\phi, kick}, v_{z, kick})$ was generated for each MSP. A hard cutoff of 600 km/s was enforced for any given velocity component, as MSPs with velocities exceeding 600 km/s would most likely be ejected from the Milky Way Galactic 
Bulge 
and not contribute to the GCE. Fig.~\ref{fig:maxwellian_kicks} 
illustrates 
the two Maxwellian natal kick distributions.

Our next step was to place our MSPs into the Galaxy. To do this, we 
used 
the total stellar mass contained within each Milky Way mass component
\citep{Bland-Hawthorn16}, 
as
listed in Table~\ref{tab:stellar_masses}.
\begin{table}
	\centering
	\begin{tabular}{cc} 
		\hline
		Galactic Component & Total Stellar Mass [$M_\odot$]\\
		\hline
		Galactic Disk (GD) & $^* 4.1 \times 10^{10}$\\
		Nuclear Stellar Cluster (NSC) & $1.8 \times 10^{7}$\\
		Galactic Bulge (GB) & $1.7 \times 10^{10}$\\
		Nuclear Stellar Disk (NSD) & $1.4 \times 10^{9}$\\
		Halo & $5.5 \times 10^{8}$\\
		\hline
	\end{tabular}
	\caption{
	Estimated stellar mass contained within each Galactic component,
	following 
	\citet{Bland-Hawthorn16}.
	$^*$ Both the stellar mass of the thin ($3.5 \times 10^{10} M_\odot$) and thick ($6 \times 10^{9} M_\odot$) disk were incorporated into the total stellar mass of the GD.}
	\label{tab:stellar_masses}
\end{table}

Initial cylindrical Galactocentric coordinates were generated for all LMXBs/MSPs from stellar mass density priors. Although it is not the only formation pathway, the standard formation mechanism assumes LMXBs and thus MSPs are produced from stellar binaries formed within the various stellar mass components of the Galaxy. 
We expect the spatial distribution of LMXBs/MSPs to, initially, follow the stellar mass profiles in our Galaxy.
\citet{Voss07b} argue that LMXBs in the central bulge of M31 (the central 1', corresponding to 1.6 degrees for the Milky Way bulge) are dominated by systems formed through dynamical interactions due to the relatively high density of the bulge. However, their calculations of dynamical interactions at bulge velocity dispersions find 5 times more LMXBs produced with black holes than with neutron stars; such systems cannot produce MSPs, and thus this route looks less promising to produce the GCE. Later we will consider alternative scenarios in which MSPs are deposited from destroyed globular clusters \citep{Brandt15}.

Using the stellar mass profiles for the GD, the NSC, the NSD, the GB, and the Halo
\citep{Launhardt02,Bland-Hawthorn16,Xue15},
we normalized these profiles with the total masses 
\citep{Bland-Hawthorn16} 
contained within each component. For the GD, we utilized the standard double-exponential mass density profile with its accepted scale length parameters (\ref{eq:gd}) 
\begin{equation}
    \rho(r,z)=\rho_0e^{\frac{-r}{h_r}}e^{\frac{-|z|}{h_z}}
	\label{eq:gd}
\end{equation}
where $h_r$ = 4 kpc and $h_z$ = 0.1 kpc. For the NSC, the profile we used follows (\ref{eq:nsc}) 
\citep{Launhardt02}: 
\begin{equation}
    \rho(r,z) =
    \begin{cases} 
      \frac{\rho_0}{1+(\frac{\sqrt{r^2+z^2}}{h_r})^n} & \sqrt{r^2+z^2} < 1.2\;kpc \\
      0 & \sqrt{r^2+z^2} \geq 1.2\;kpc \\
   \end{cases}
   \label{eq:nsc}
\end{equation}
where n = 2 and $h_r$ = $2.2\times10^{-4}$ kpc. The piecewise nature of the NSC profile was adopted such that the binaries born within the NSC remained concentrated around the GC. Our NSD stellar mass profile follows (\ref{eq:nsd})
\citep{Launhardt02}:
\begin{equation}
    \rho(r,z) = \rho_0e^{ln(0.5)|\frac{2r}{h_r}|^{n_r}}e^{ln(0.5)|\frac{2z}{h_z}|^{n_z}}
   \label{eq:nsd}
\end{equation}
where $n_r$ = 5, $h_r$ = 0.17 kpc, $n_z$ = 1.4, $h_z$ = 0.045 kpc. More specifically, 
\citet{Launhardt02} 
supplied the radial component of the NSD profile above, and we made an additional assumption that the z-component of the profile would fall off in a similar manner in order to reproduce a profile similar to the standard disk double exponential profile. The stellar mass profile for the GB required more careful consideration as the GB or the `bar' of the Milky Way is believed to be rotated with respect to the Galactocentric frame. We made use of (\ref{eq:gb}) 
\citep{Bland-Hawthorn16} 
for our GB stellar mass profile
\begin{equation}
    \rho(x',y',z') = \rho_0e^{\frac{-|x'|}{h_x}}e^{\frac{-|y'|}{h_y}}e^{\frac{-|z'|}{h_z}}
   \label{eq:gb}
\end{equation}
where $h_x$ = 0.70 kpc, $h_y$ = 0.44 kpc, and $h_z$ = 0.18 kpc. Here, $(x',y',z')$ are a set of `Bulge' coordinates corresponding to cartesian Galactocentric coordinates $(x,y,z)$ rotated 27$^{\circ}$ degrees clockwise about the Galactocentric z-axis. Finally, we chose our Halo stellar mass profile to obey (\ref{eq:halo}) 
\citep{Xue15}: 
\begin{equation}
    \rho(r,z) = \rho_0(\sqrt{r^2+z^2})^{-\alpha}
   \label{eq:halo}
\end{equation}
where $\alpha$ = 2.1. We only selected the inner portion of 
\citet{Xue15}'s broken power-law profile 
since we are only studying the inner regions of the Galaxy. We set our MSP population size to have 1,000,000 MSPs distributed initially in the various stellar mass components of the Galaxy. The sizes of sub-populations in the various stellar mass components were set by the stellar mass component's contribution to the total mass contained within the central 20 kpc of the Milky Way. For example, the GD holds $\sim65\%$ of the stellar mass with the central 20 kpc, so we would sample $\sim650,000$ MSP initial coordinates from the GD profiles. Our fixation with a 20 kpc sphere is from another cutoff we enforced where we would disregard any MSPs initially placed exterior to this 20 kpc boundary, as these MSPs would have little influence on the GCE. To summarize, with the normalized stellar mass profiles above, we would sample an appropriate number of initial coordinates from each stellar component proportional to the component's contribution to the total mass enclosed within the Milky Way.

\begin{figure}
	\includegraphics[width=\columnwidth]{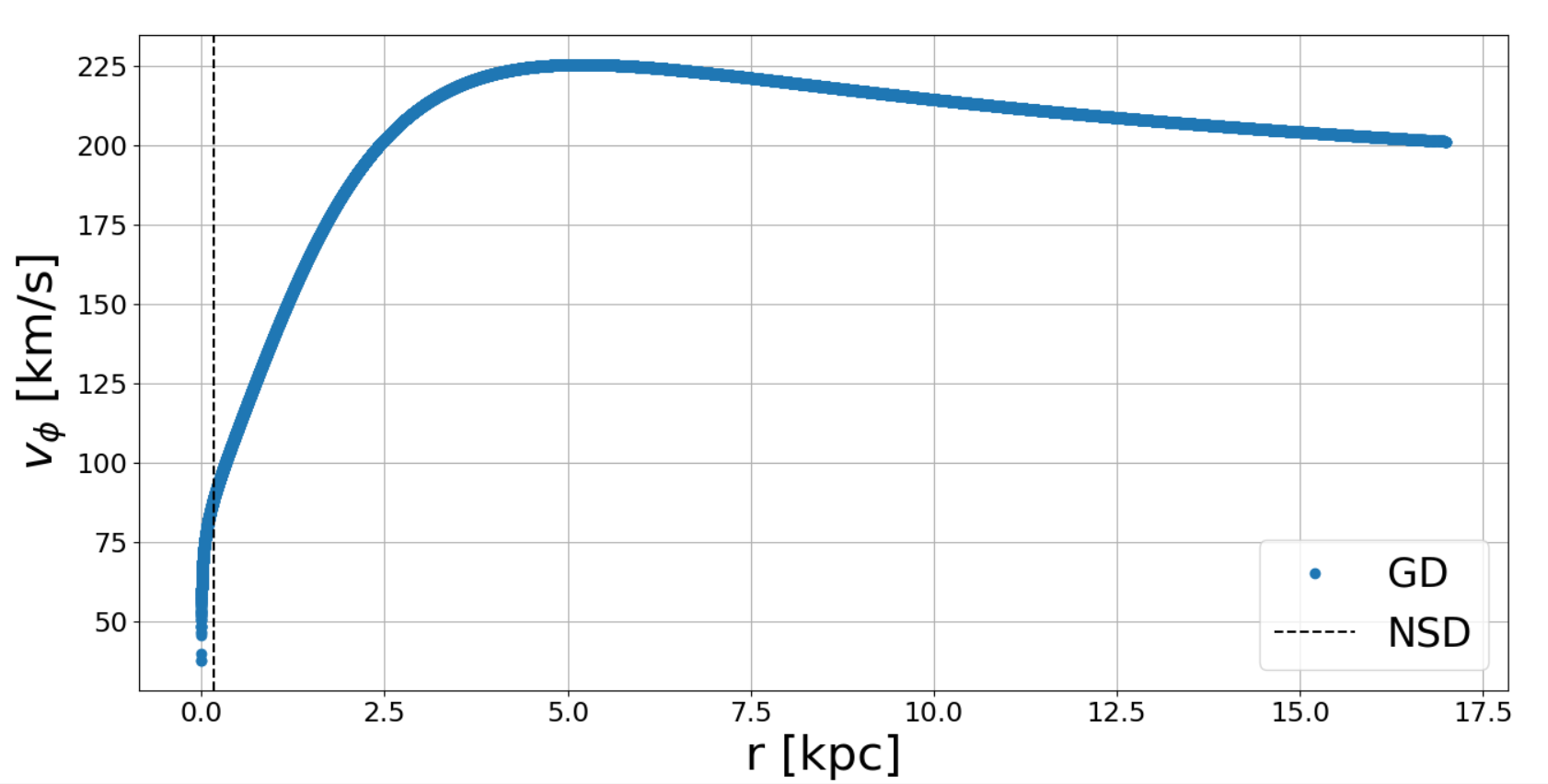}
    \caption{
    Intrinsic orbital, tangential velocities supplied to MSPs born either in the GD or NSD. The tangential velocities of the GD-born MSPs are shown in blue as a function of cylindrical, Galactocentric radius. The intersection between the black, dashed line and blue curve shows the tangential velocities given to NSD-born MSPs.}
    \label{fig:rotation_curves}
\end{figure}

We adopted a relatively standard prescription for assigning orbital velocities to each LMXB intrinsic to their birthplace component. Disk-like components such the GD or the NSD provided purely tangential velocities from a rotation curve generated, using \textit{galpy}'s functionality, from our realistic Milky Way potential. Less structured components like the Halo, the NSC, or the GB would supply their binaries with uniform magnitude, randomly oriented velocities compatible with observations
\citep{Schodel09,Valenti18}. 
In more detail, $\vec{v}_{NSC} = \sqrt{3}\sigma_{NSC}\vec{u}$, $\vec{v}_{GB} = \sqrt{3}\sigma_{GB}\vec{u}$, $\vec{v}_{Halo} = 220\vec{u}$ where 1D velocity dispersions are $\sigma_{NSC}$ = 100 km/s and $\sigma_{GB}$ = 140 km/s 
\citep{Schodel09,Valenti18}, 
and $\vec{u}$ is a randomly oriented unit vector. For the GD and NSD, $v_r$, $v_z$ = 0. Intrinsic GD $v_{\phi}(r)$ were drawn directly from our Milky Way potential rotation curve whereas all intrinsic NSD $v_{\phi}(r)$ were set to $v_{\phi}(0.17\;kpc)$ = 88.7 km/s. Fig.~\ref{fig:rotation_curves} depicts the intrinsic rotation curves applied to the MSPs of disk-like stellar mass components.
The total, initial velocity of a given MSP was set as the superposition between these velocities provided intrinsically from the stellar mass components and the sampled natal kick velocities. So, our initial parameters could be divided into three sub-groups dependent on the kick prescription where 1,000,000 MSPs were simulated for each kind of kick.

With our initial phase space elements obtained, we then had to set the gravitational environment of which our MSP population would traverse through. To do this, we had to implement and craft a realistic gravitational potential mimicking the one of the Milky Way. Making use of \textit{galpy}'s built-in potential class, we introduced the \textit{MWPotential2014} potential as the foundation for our gravitational potential because it produced a reasonable fit to a large variety of observed Milky Way data
\citep{Bovy15}. 
\textit{MWPotential2014} models the GB potential using a \textit{PowerSphericalPotentialwCutoff} potential, the GD potential with a \textit{MiyamotoNagaiPotential}, and the dark matter Halo potential using a \textit{NFWPotential}. However, for our purposes, we needed to include the gravitational effects of the NSC along with the supermassive black hole Sgr A* as these components play a strong role in the dynamical evolution of stellar populations near the GC. We superposed a \textit{PlummerPotential} 
\citep{Pflamm-Altenburg09}, 
to model the NSC, and a \textit{KeplerPotential}, to model Sgr A*, with the \textit{MWPotential2014} to produce the final gravitational potential acting on our MSP population.

For each population of 1,000,000 MSPs, characterized by their unique natal kick prescription, we would evolve them in this realistic Milky Way gravitational potential for 1 Gyr. This simulation length was chosen such that the MSPs could settle into a final, stable configuration near the GC or be ejected from the Galaxy altogether. Every 100 Myrs, the cylindrical Galactocentric coordinates of each MSP would be collected. Although the dynamical trajectory holds rich information itself, we were particularly interested in the final locations of each MSP which we treated as an analog of today's MSP distribution.

\section{Results and Discussion}

\subsection{21 degree analysis}
Previous studies have supplied detailed observational data of the GCE
\citep{diMauro21,Horiuchi16,Daylan16}, 
which we made use of in our comparison between simulated and observed data. We believe the differences in GeV emission in the central regions of the Galaxy for \citet{diMauro21}, \citet{Horiuchi16}, and \citet{Daylan16} arise from their different filtering analysis, including interstellar emission models. The observational data was provided as surface brightness as a function of angular radius seen on the sky. So, we had to craft our own surface brightnesses for each annulus. To do this, we extracted the final cylindrical coordinates, for each 1,000,000 MSP population, to generate a 
2D
spatial distribution of MSPs. Using \textit{astropy}'s \citep{Astropy2018} coordinate transformation capabilities, we were able to transform our Galactocentric coordinates $(r,\phi,z)$ into Galactic coordinates $(l,b,s)$ where $s$ is the distance to the source from the Sun. At this point, we defined the angular radius of a given MSP to be $\sqrt{l^2+b^2}$.

\begin{figure}
	\includegraphics[width=\columnwidth]{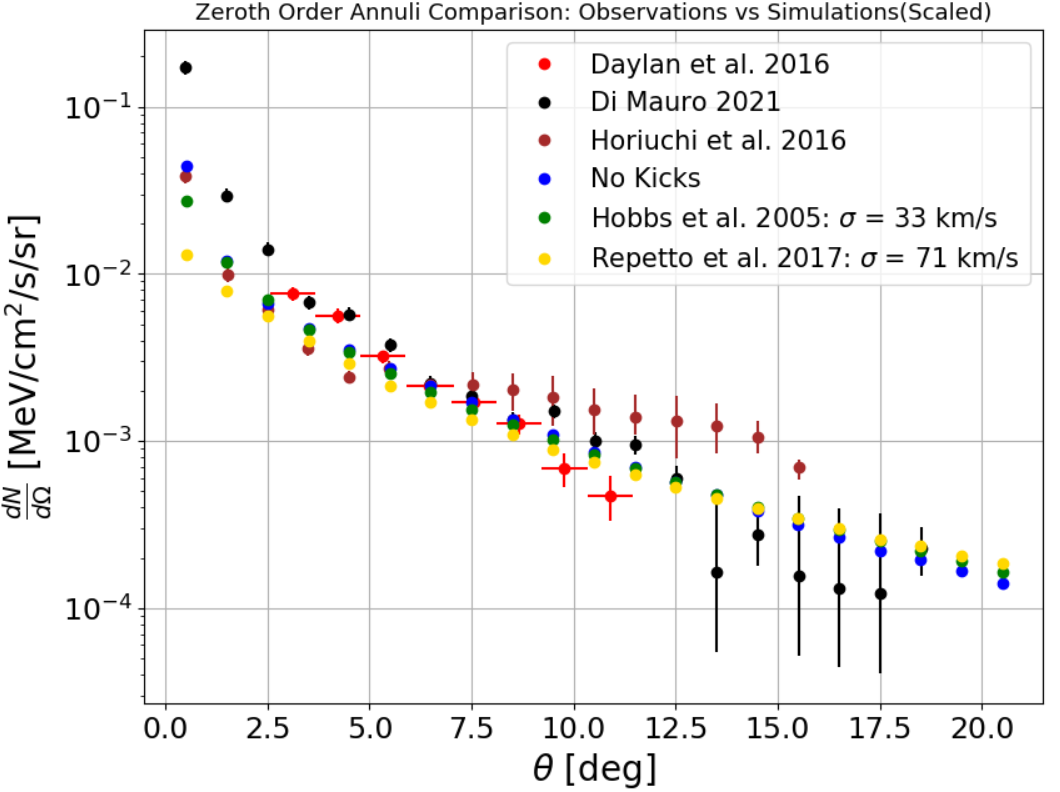}
    \caption{
    Observed surface brightnesses (from Daylan, Di Mauro, and Horiuchi; Daylan omits the central few degrees)
    compared with our 
    synthetic distributions of MSPs (with different initial kick velocities), as a function of angle from the Galactic Centre. 
    In this zeroth order approximation, all MSPs in our simulations reside 
    8.2 kpc away from us, and, hence, surface brightnesses in each annulus are directly proportional to the number of gamma-ray producing MSPs in each annulus. Differences can be seen among the observed surface brightnesses, but clearly the observed surface brightnesses tend to be steeper in the central few degrees than MSP distributions with realistic kick prescriptions. 
    }
    \label{fig:zeroth_order}
\end{figure}

\begin{figure}
	\includegraphics[width=\columnwidth]{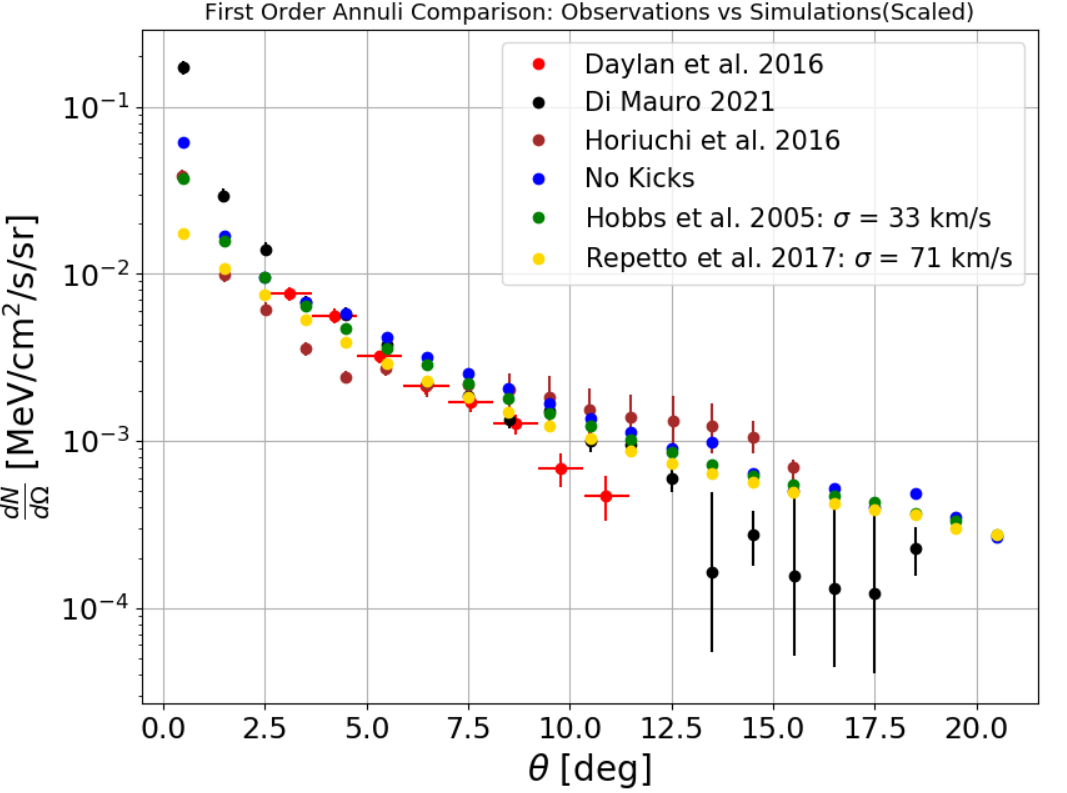}
    \caption{
    Observed surface brightnesses compared to synthetic surface brightnesses of each annulus across the sky. Under this first order approximation, 
    we use the inferred distance of each MSP in our simulation, 
    and, hence, surface brightnesses in each annulus are directly proportional to the fluxes of gamma-ray producing MSPs in each annulus. The differences with Fig.~\ref{fig:zeroth_order} are subtle.}
    \label{fig:first_order}
\end{figure}

We computed surface brightness using two orders of approximation. To zeroth order, all MSPs in the GC are roughly the same distance from the Earth, and, hence, the surface brightness in a given annulus on the sky is directly proportional to the number of MSPs within that annulus. So we would count up the number of MSPs within a range of angular radii, divide by the solid angle of the annulus, and report the corresponding approximate surface brightness. Fig.~\ref{fig:zeroth_order} illustrates the zeroth order comparison between our data and the observed data sets. For a first order approximation, we consulted the literature for a MSP gamma-ray luminosity, L$_{\gamma,MSP}$ = $4.7\times10^{25}$ W 
\citep{Hooper16},  
and computed gamma-ray fluxes of each annulus using the distances along line of sight. Fig.~\ref{fig:first_order} showcases this higher order comparison. We scaled our synthetic profiles to match the overall normalization of the  observed GeV profiles provided in \citet{diMauro21}, \citet{Horiuchi16}, and \citet{Daylan16}, allowing comparison of the shapes of the curves  over the angular range. This is justified, as we are uncertain about the number of pulsars residing in the central parts of the Galaxy, and their corresponding gamma-ray luminosity, which introduces an arbitrary normalization factor.

We fit, using least squares, our simulated data to constrain this parameter for each supplied data set and each type of kick (9 in total). 
We report 
the total luminosity produced by all the simulated MSPs within our area of interest (central $21^{\circ}$ on the sky) as this result would be independent of population size once fit to our observed data. 
Table~\ref{tab:luminosities} lists our results.
\begin{table}
	\centering
	\begin{tabular}{cccc} 
		\hline
		Natal Kick($\sigma$) & Di Mauro 2021 & Horiuchi 2016 & Daylan 2016\\
		\hline
		No Kicks & $6.96\pm0.14$ & $4.48\pm0.09$ & $5.00\pm0.11$\\
		33$\frac{km}{s}$ & $7.65\pm0.15$ & $4.93\pm0.10$ & $5.38\pm0.12$\\
		71$\frac{km}{s}$ & $8.41\pm0.17$ & $5.45\pm0.11$ & $5.57\pm0.12$\\
		\hline
	\end{tabular}
	\caption{
	Total gamma-ray luminosities ($\times10^{29}$ W) produced from our simulated MSPs in the central $21^{\circ}$ for each observed data set and kick prescription. Our `No Kicks' are simulations where LMXBs/MSPs were not provided any natal kicks and ended with 403,266 MSPs within the central $21^{\circ}$.  33$\frac{km}{s}$ simulations use the kick prescription in 
	\citet{Hobbs05} 
	and had 382,490 MSPs within the central $21^{\circ}$. 71$\frac{km}{s}$ simulations use the kick prescription in 
	\citet{Repetto17} 
	and had 324,278 MSPs within the central $21^{\circ}$. We simply fit our simulated surface brightnesses to our observed data sets to determine the gamma-ray luminosity required for a single MSP and multiplied this parameter by the total MSPs within the central $21^{\circ}$, effectively removing the dependence on the number of MSPs simulated.}
	\label{tab:luminosities}
\end{table}
With our fit luminosities, we then linearly interpolated our simulated first order surface brightnesses in order to make an adequate comparison between the three observed data sets. A logarithmic $\chi^2$ analysis was performed for each type of kick for each data set where the difference in the logarithms of the simulated data and observations determined the $\chi^2$ statistic. When errorbars were provided, 
we used a logarithmic average of the upper and lower error bars. Observed data points without error bars were given a 10$\%$ error, and observed upper limits were scaled down by a factor of 3 and provided with error bars such that the upper error bar reached the original, unscaled upper limit. Table~\ref{tab:log_chi} demonstrates the results from our $\chi^2$ analysis, where the number of degrees of freedom were 18 for comparisons with
\citet{diMauro21}, 
15 for comparisons with
\citet{Horiuchi16},
and 7 for comparisons with 
\citet{Daylan16}.
Since the Fermi-LAT data collected from 
\citet{diMauro21} 
spans the largest angular range across the sky, providing the most stringent constraints, we are inclined to interpret comparisons with this data set with more weight than the others. 
\begin{table}
	\centering
	\begin{tabular}{cccc} 
		\hline
		Natal Kick($\sigma$) & Di Mauro 2021 & Horiuchi 2016 & Daylan 2016\\
		\hline
		No Kicks & 235.290 & 125.647 & 30.683\\
		33$\frac{km}{s}$ & 324.931 & 114.323 & 29.066\\
		71$\frac{km}{s}$ & 554.987 & 165.682 & 29.304\\
		\hline
	\end{tabular}
		\caption{
		$\chi^2$ values for the fit with each kick prescription to the three observed data sets, using the full angular range. 
	There were 18 degrees of freedom in the analysis using  
	\citet{diMauro21},
	15 in the analysis using  
	\citet{Horiuchi16},
	and 7 in the analysis using  
	\citet{Daylan16}. There is a strong preference for No Kicks from the 
	\citet{diMauro21} data, although even this fit is not very good. As the \citet{Daylan16} data omits the central few degrees, it is much less sensitive to the variation in initial kicks. }
	\label{tab:log_chi}
\end{table}

As seen in this table, there is a relatively strong preference for pulsars without kicks. It is clear from Fig.~\ref{fig:first_order} that 
the observed data differ significantly from the simulated data. 
The simulated MSP populations cannot generate surface brightnesses high enough relative to the 
\citet{diMauro21} profile 
in the central 
three degrees 
of the Galaxy. Even MSPs without kicks, the most concentrated population, cannot produce such high surface brightnesses. 
It thus seems unlikely that gamma-rays produced by primordially formed MSPs can be  the process producing the bulk of the GCE.

\subsection{5 degree analysis}
We also tried an analysis focusing only on the central 5 degrees around the Galactic Center. 
For this analysis, we do not use data from 
\citet{Daylan16}, 
which does not cover this region. 
For $\theta$ > 5$^{\circ}$, other gamma-ray sources apart from emission from the GC 
could influence 
the gamma-ray observations made by 
\citet{diMauro21,Horiuchi16,Daylan16}. 
These studies would have had to filter out these intruding signals, which introduce another layer of uncertainty in their reported surface brightnesses at larger angular radii. Hence, smaller angular radii data may be more accurate. 

Since we are examining the difference in profile behaviour and not the magnitudes of the brightnesses, we were free to introduce arbitrary normalizations to get the simulated and observed data as close as possible before performing another logarithmic $\chi^2$ analysis. We performed a similar type of normalization as in the central 21 degrees analysis where we normalized our synthetic profiles to the observed data in the central 5 degrees of the Galactic Center. Scaling all three simulated data sets to both the surface brightness profile of 
\citet{diMauro21} 
and the surface brightness profile of 
\citet{Horiuchi16}, 
we calculated the logarithmic $\chi^2$ with 4 degrees of freedom for both observed data sets as shown in Table~\ref{tab:central_chi}.
\begin{table}
	\centering
	\begin{tabular}{ccc} 
		\hline
		Natal Kick($\sigma$) & Di Mauro 2021 & Horiuchi 2016\\
		\hline
		No Kicks & 83.781 & 19.178\\
		33$\frac{km}{s}$ & 151.437 & 35.480\\
		71$\frac{km}{s}$ & 298.930 & 122.991\\
		\hline
	\end{tabular}
	\caption{
	$\chi^2$ values for comparison of simulated data normalized to the observed data, both with 4 degrees of freedom, in the central 5 degrees on the sky. 
	The normalization ensures that the shapes are compared. 
	It is clear that none of the simulated populations produce a gamma-ray emission profile very consistent with the GCE. If primordial MSPs produce all the GCE, this table indicates that low-kick MSPs are 
	strongly preferred.}
	\label{tab:central_chi}
\end{table}

\begin{figure}
	\includegraphics[width=\columnwidth]{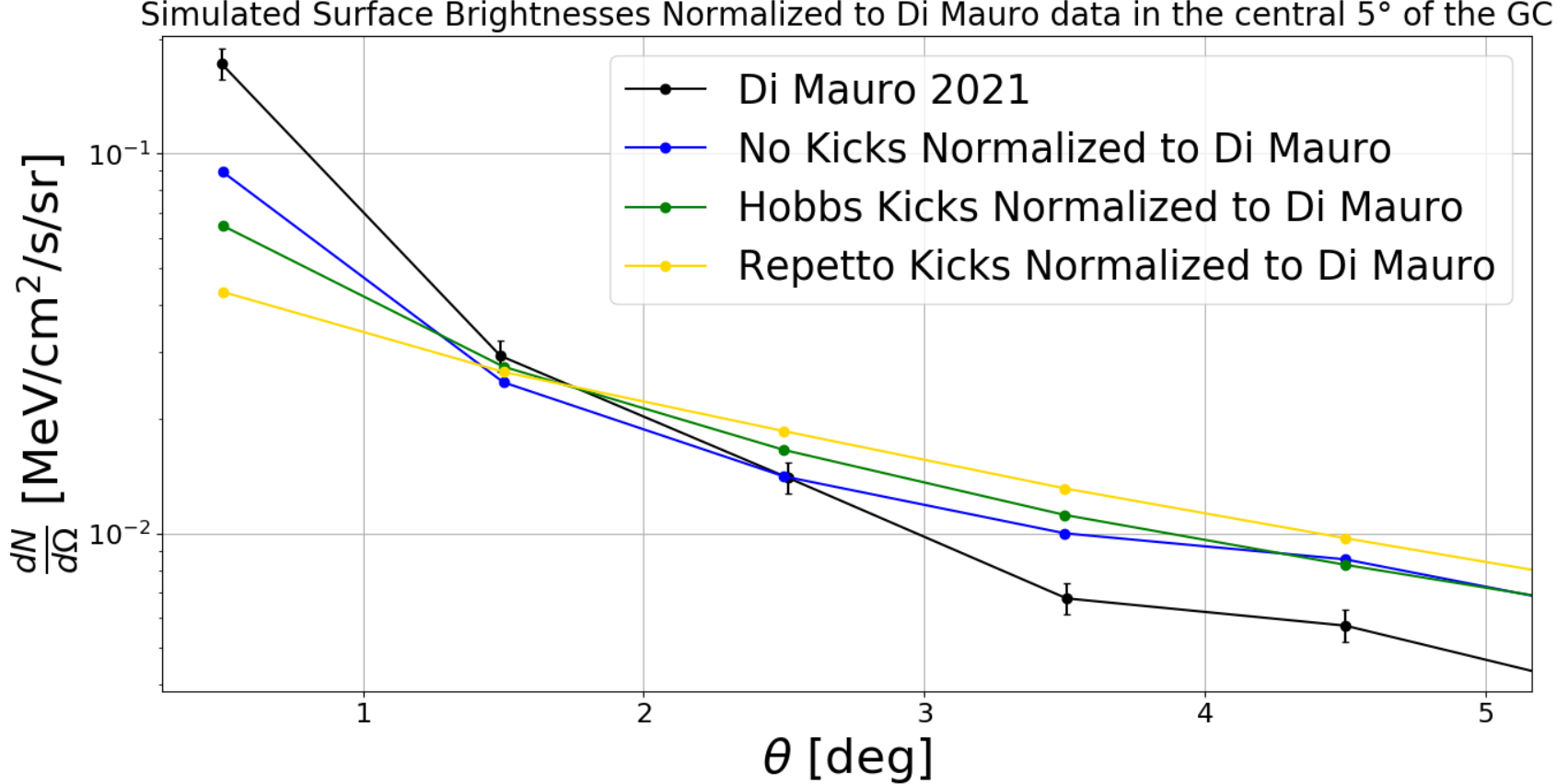}
    \caption{
    Observed surface brightnesses from 
    \citet{diMauro21} 
    plotted against synthetic surface brightnesses from each kick prescription in the central 
    5 degrees. 
    The lines connecting all data points are from linear interpolation. The simulated data do not appear to match the shape of the observed brightness profile.}
    \label{fig:central_fermi}
\end{figure}

\begin{figure}
	\includegraphics[width=\columnwidth]{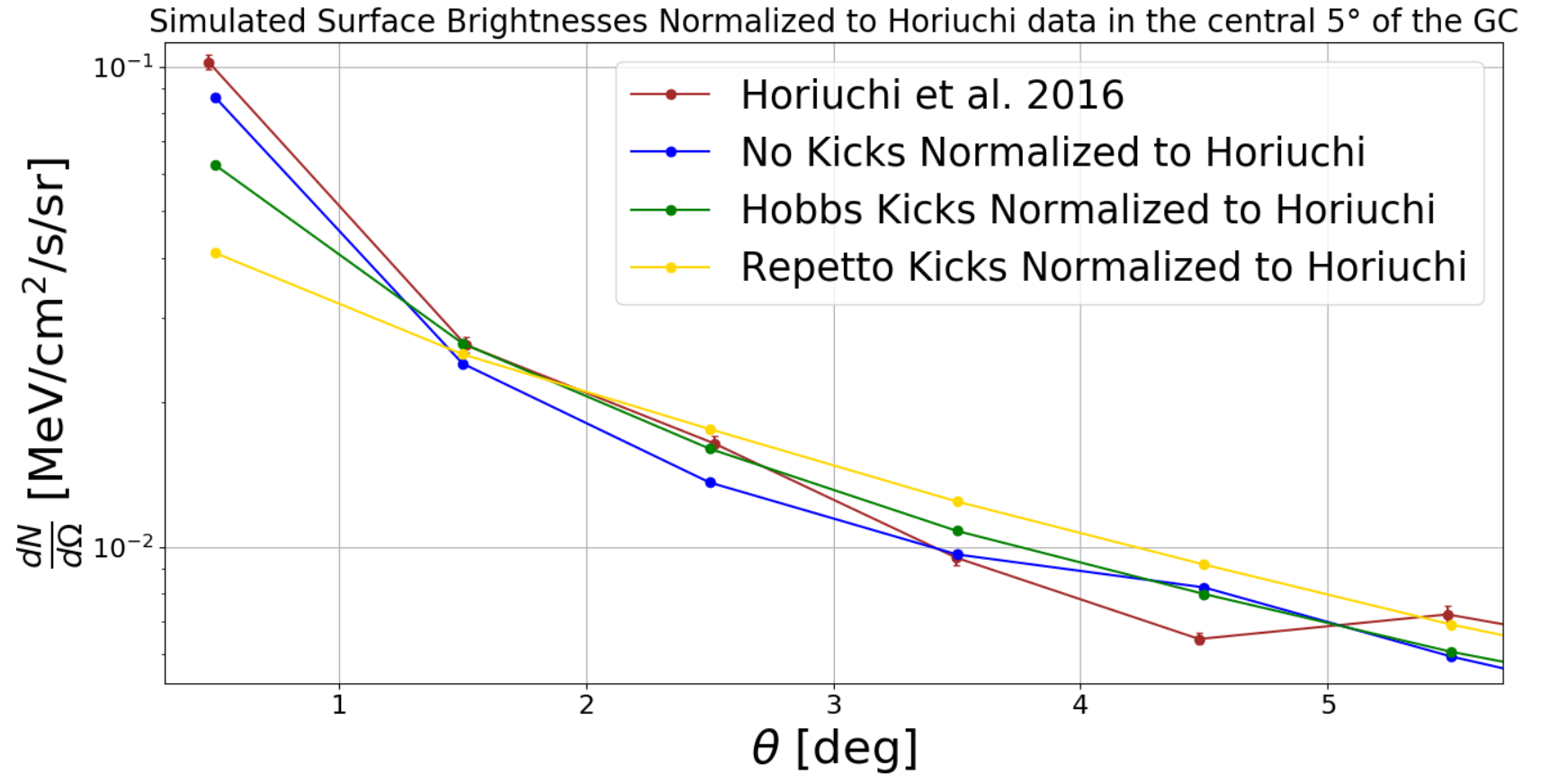}
    \caption{
    Observed surface brightnesses from 
    \citet{Horiuchi16} 
    plotted against synthetic surface brightnesses from each kick prescription in the central 
    5 degrees. 
    The lines connecting all data points are from linear interpolation. The population of MSPs without natal kicks and the observed data here agree relatively well, implying that a population of MSPs without kicks, residing in the GC, could produce the GCE.}
    \label{fig:central_horiuchi}
\end{figure}

Fig.~\ref{fig:central_fermi} and Fig.~\ref{fig:central_horiuchi} contrast the varying surface brightness profiles between observed and simulated data.
Our results are essentially similar to those using the full 21 degree datasets; the Di Mauro dataset is too centrally concentrated to be explained by any of our simulated data, while the Horiuchi data can be reasonably fit only by our simulation involving no initial neutron star kicks, which 
does not appear 
physically plausible. 
These results imply either: (i) MSPs are not the origin of the GCE; or (ii) MSPs 
in the GC are formed and deposited via other processes, such as dynamical formation mechanisms. 

\subsection{Inspiraling globular clusters}
With the capabilities of \textit{galpy} and our simulated data, we were able to investigate the plausibility of globular cluster deposited MSPs as an alternative/additional source of MSPs in the GC. 
Globular clusters that formed 
in the 
young Milky Way may have inspiralled to the GC due to dynamical friction 
\citep{Gnedin14}.
Once tidally disrupted and amalgamated into the Milky Way, the MSPs within such clusters may have been deposited into the GC, resulting in an excess relative to the population of MSPs born from our Galaxy's stellar mass components. Hence, low natal kick MSPs from globular clusters could potentially contribute to the gamma-ray excess. There are two primary questions one needs to address when looking at this mechanism: 
(1) How many MSPs are required, in excess of the Galaxy-born MSPs, to boost gamma-ray surface brightnesses in the central regions of the sky to agree with observational data? (2) What is the expected radial profile of globular cluster debris?

To address (1), our analysis boils down to determining the extra MSPs required to inflate the central regions of our surface brightness profiles, seen in Fig. ~\ref{fig:first_order}, to match the \citet{diMauro21} 1-10 GeV data, along with the data of 
\citet{Horiuchi16}.
Assuming our simulated and observed brightness profiles are in agreement for $\theta \geq 5^\circ$, we normalized our simulated surface brightness for $\theta \epsilon [3^\circ,5^\circ]$ to ensure the difference in brightness, between simulated and observed profiles, represented extra luminosity that needed to be injected via globular cluster deposited MSPs. This normalization ensures the difference between profiles is independent of simulated population size. With the difference in brightnesses and annulus solid angles on hand, we were able to compute the excess flux required for agreement with observations. Here, we make a zeroth order approximation that all simulated MSPs within the central region of the sky (central $3.5^\circ$) are MSPs residing in close proximity to the GC. Hence, all extra MSPs are approximately 8.2 kpc away from us. With this approximation, we can then compute the extra luminosity in excess of the gamma-ray luminosity provided by primordial Galaxy-born MSPs alone. Table~\ref{tab:extra_lum} depicts our analysis.
\begin{table}
	\centering
	\begin{tabular}{ccc} 
		\hline
		Natal Kick($\sigma$) & Di Mauro 2021 & Horiuchi 2016\\
		\hline
		No Kicks & 2.180 & 0.2455\\
		33$\frac{km}{s}$ & 2.236 & 0.2604\\
		71$\frac{km}{s}$ & 2.587 & 0.4436\\
		\hline
	\end{tabular}
	\caption{
	Extra gamma-ray luminosity ($\times10^{29}$ W) needed to inflate 
	the central part of 
	our synthetic surface brightnesses to match the sharp increase seen in the observed data. These excess luminosities can correct our simulated profiles such that we have agreement in the central regions of the sky.}
	\label{tab:extra_lum}
\end{table}

Making use of an average MSP gamma-ray luminosity L$_{\gamma,MSP}$ = $4.7\times10^{25}$ W
\citep{Hooper16},
we can provide an estimate of the number of MSPs required from inspiralling globular clusters. If globular cluster deposited MSPs were the source of this excess luminosity, approximately 4,966 MSPs or 673 MSPs would have to be injected, according to 
\citet{diMauro21}, 
 or 
\citet{Horiuchi16},
respectively. 
It has been estimated that of order 1000-2000 MSPs lie within globular clusters orbiting the Milky Way today 
\citep[][Zhao \& Heinke,  submitted]{Heinke05,Turk13}. 
Thus, the larger population size estimates would require substantially more (a factor of 5) globular clusters to be destroyed in the inner Milky Way, than exist today.\footnote{The relevant factor is actually stellar encounter rate, for MSP production; it is unclear if destroyed globular clusters would be preferentially of higher or lower encounter rates than surviving ones.} Some studies of the destruction of globular clusters do predict ratios of this order \citep{Horta21}.
We therefore consider adding together our simulated primordial MSP profile with  dynamically deposited MSP profiles (below).

\subsection{Nuclear stellar cluster}

It has been postulated that, due to its similarities with a globular cluster dynamical environment, the NSC would have enhanced MSP produced rates from dynamical captures in addition to in situ formation.
\citet{Muno05} observed several X-ray transients (likely LMXBs) within the central parsec (0.45') around Sgr A*, and performed a toy calculation estimating that 100-1000 neutron stars and black holes in the NSC could have exchanged into binaries.
\citet{Generozov18} performed detailed Fokker-Planck calculations of nuclear star clusters, and estimated the rates of tidal captures, predicting of order 100 LMXBs containing neutron stars, and another 100 containing black holes, within 3.5 pc of Sgr A*. Assuming typical lifetimes of 1 Gyr for LMXBs and 10 Gyrs for MSPs, this would suggest of order 1000 MSPs produced in the NSC.  
\citet{Abbate18} calculated the encounter rate for the NSC, finding a stellar encounter rate similar to that of Terzan 5 \citep{Bahramian13}, which is thought to hold of order 150 MSPs \citep{Bagchi11};  \citet{Abbate18} prefer to explain MSPs in the central parsecs as deposited by globular clusters inspiraling and dissolving. On the other hand, \citet{Faucher-Giguere11} proposed factors that would increase the number of MSPs produced in the NSC, preferring an estimate of 500 MSPs for the NSC. We do not attempt our own theoretical calculation of the NSC population of MSPs, but below we fit the observed GCE with a model including an NSC population (which is a point source at our resolution), and then will compare the required number to the estimates above.


\subsection{Multiple mechanisms}
Finally, we implemented another approach to assess whether or not a primordial (Galaxy-born) MSP population coupled with a globular cluster deposited MSP population and a dynamically formed NSC MSP population could 
explain 
the GCE. 
With this approach, we could estimate the relative contributions of each of 
these 
MSP populations to the GCE. 
To predict the distribution of MSPs from globular cluster dynamical evolution, we used the detailed simulations of \citet{Fragione18}. Making use of \citet{Fragione18}'s Figure 4, we implemented both their Model GAU-K14C and Model LON-K14 as potential models of gamma-ray surface brightness emission from globular cluster deposited MSPs. Further, we also utilize the data shown on \citet{Fragione18}'s Figure 1 which implements models from \citet{Gnedin14}, and calculate the surface brightness profile expected from such a cumulative stellar mass distribution. In the end, we 
fit the following overall surface brightness incorporating all three MSP sources:
\begin{equation*}
    \frac{dN}{d\Omega} =
    \alpha \frac{dN}{d\Omega}_{\alpha} + \beta \frac{dN}{d\Omega}_{\beta}
    + \gamma
    , \ \gamma \ \epsilon \ \mathbb{R}
   \label{eq:fit}
\end{equation*}
where
\begin{equation*}
    \gamma = 
    \begin{cases}
      \gamma & \theta < 0.6^\circ \\
      0 & \theta \geq 0.6^\circ \\
   \end{cases}
   \label{eq:gamma}
\end{equation*}
For the subscripts, $\alpha$ indicates the primordial MSP contribution from our N-body simulations, $\beta$ indicates the globular cluster MSP contribution from \citet{Fragione18}, and $\gamma$ indicates an unknown, constant NSC MSP contribution. The piecewise nature of $\gamma$ follows from how dynamically formed MSPs in the NSC region of the Galaxy 
are constrained only to the NSC itself. 
We fit for $\alpha$, $\beta$, and $\gamma$ using the 
radial distributions 
of \citet{diMauro21} and \citet{Horiuchi16} contained in the central 3.5$^\circ$. Table~\ref{tab:fit_param} highlights our best fit parameter values, assessed from their $\chi^2$ statistic.
\begin{table}
  \begin{tabular}{|l|l|l|l|l|l|l|}
    \hline
    \hline
    \multirow{3}{*}{ Model GAU-K14C} &
      \multicolumn{1}{c}{\textbf{No Kicks}} \\
    & Di Mauro 2021 & Horiuchi et al. 2016  \\
    \hline
    $N_{\alpha}$ & $0\substack{+4871 \\ -0}$ & $527\substack{+3007 \\ -527}$ \\
    $N_{\beta}$ & $5956\pm3980$ & $2283\pm1574$ \\
    $N_{\gamma}$ & $164\pm26$ & $15\pm7$ \\
       $\chi^2$/dof & $8.982$ & $1.418$\\
    \hline
    \multirow{1}{*}{ Model LON-K14} \\
    \hline
    $N_\alpha$ & $0\substack{+9242 \\ -0}$ & $0\substack{+6158 \\ -0}$ \\
    $N_\beta$ & $4998\substack{+6315 \\ -4998}$ & $2146\substack{+2696 \\ -2146}$ \\
    $N_\gamma$ & $179\pm59$ & $23\substack{+25 \\ -23}$ \\
       $\chi^2$/dof & $13.835$ & $1.005$\\
    \hline
    \multirow{1}{*}{ Gnedin et al. 2014} \\
    \hline
    $N_\alpha$ & $7322\pm5858$ & $4904\pm2813$ \\
    $N_\beta$ & $0\substack{+6021 \\ -0}$ & $0\substack{+1885 \\ -0}$ \\
    $N_\gamma$ & $136\pm58$ & $4\substack{+8 \\ -4}$ \\
       $\chi^2$/dof & $16.352$ & $2.417$\\
    \hline
    \hline
    \multirow{3}{*}{ Model GAU-K14C} &
      \multicolumn{1}{c}{\textbf{Hobbs et al. 2005 Kicks}} \\
    & Di Mauro 2021 & Horiuchi et al. 2016  \\
    \hline
    $N_\alpha$ & $0\substack{+3712 \\ -0}$ & $2009\substack{+2408 \\ -2009}$ \\
    $N\beta$ & $5956\pm3382$ & $1387\substack{+1406 \\ -1387}$ \\
    $N_\gamma$ & $164\pm20$ & $20\pm6$ \\
    $\chi^2$/dof & $8.982$ & $1.1$\\
    \hline
    \multirow{1}{*}{ Model LON-K14} \\
    \hline
    $N_\alpha$ & $0\substack{+10705 \\ -0}$ & $0\substack{+6520 \\ -0}$ \\
    $N_\beta$ & $4998\substack{+8142 \\ -4998}$ & $2146\substack{+3180 \\ -2146}$ \\
    $N_\gamma$ & $179\pm19$ & $23\pm5$ \\
      $\chi^2$/dof & $13.835$ & $1.005$\\
    \hline
    \multirow{1}{*}{ Gnedin et al. 2014} \\
    \hline
    $N_\alpha$ & $6573\substack{+7223 \\ -6573}$ & $4389\pm3041$ \\
    $N_\beta$ & $0\substack{+8273 \\ -0}$ & $0\substack{+2280 \\ -0}$ \\
    $N_\gamma$ & $181\pm68$ & $24\pm20$ \\
      $\chi^2$/dof & $17.188$ & $1.585$\\
    \hline
    \hline
    \multirow{3}{*}{ Model GAU-K14C} &
      \multicolumn{1}{c}{\textbf{Repetto et al 2018 Kicks}} \\
    & Di Mauro 2021 & Horiuchi et al. 2016  \\
    \hline
    $N_\alpha$ & $0\substack{+1568 \\ -0}$ & $871\substack{+1046 \\ -871}$ \\
    $N_\beta$ & $5956\pm1574$ & $1999\pm674$ \\
    $N_\gamma$ & $164\pm22$ & $21\pm8$ \\
      $\chi^2$/dof & $8.982$ & $1.099$\\
    \hline
    \multirow{1}{*}{ Model LON-K14} \\
    \hline
    $N_\alpha$ & $0\substack{+2210 \\ -0}$ & $0\substack{+1484 \\ -0}$ \\
    $N_\beta$ & $4998\pm1841$ & $2146\pm801$ \\
    $N_\gamma$ & $179\pm22$ & $23\pm7$ \\
      $\chi^2$/dof & $13.835$ & $1.005$\\
    \hline
    \multirow{1}{*}{ Gnedin et al. 2014} \\
    \hline
    $N_\alpha$ & $0\substack{+4017 \\ -0}$ & $378\substack{+5790 \\ -378}$ \\
    $N_\beta$ & $7544\pm5006$ & $3009\substack{+4380 \\ -3009}$ \\
    $N_\gamma$ & $121\pm63$ & $0\substack{+54 \\ -0}$ \\
      $\chi^2$/dof & $22.817$ & $4.563$\\
    \hline
  \end{tabular}
  \caption{This table demonstrates the best fit relative contributions of MSP sources ($\alpha$ for primordial MSP contributions, $\beta$ for MSPs from destroyed globular clusters, $\gamma$ for MSPs from the NSC) to the GCE, along with the $\chi^2$ and number of degrees of freedom (dof). Each entry in the table represents the best fit number of MSPs contained in each contribution assuming a L$_{\gamma,MSP}$ = $4.7\times10^{25}$ W \citep{Hooper16} and d$_{MSP}$ = 8.2 kpc. 
	}
	\label{tab:fit_param}
\end{table}

The results of this analysis are the following. 
The majority of fits are degenerate between the contributions of the primordial MSP population and a population of MSPs from destroyed globular clusters, as long as a  nuclear star cluster contribution is included (this contribution is significant at $\gtrsim 2 \sigma$ in the majority of fits). 
However, the \citet{diMauro21} observations are significantly better fit by the GAU-K14C destroyed globular cluster model than by any of our primordial MSP models (and are better fit by any of the 3 destroyed cluster models than by primordial MSPs with Repetto kicks), while the \citet{Horiuchi16} data 
are typically degenerate between contributions from primordial MSPs vs. destroyed globular clusters, along with a small NSC contribution.
We also note that the inferred numbers of MSPs produced by the NSC in the \citet{diMauro21} fits, of order 150, are roughly consistent with the encounter rates estimated by \citet{Abbate18} for the NSC, and with the number estimated for Terzan 5. The \citet{Horiuchi16} fits require substantially smaller numbers of NSC MSPs, of order 20.

The quality of the fit parameters of Table~\ref{tab:fit_param} is also interesting.
All the fits to the \citet{diMauro21} data are significantly poorer than fits to the \citet{Horiuchi16} data; the latter are generally consistent with a "good" fit (reduced $\chi^2\sim 1$) while the former have much larger $\chi^2$ values. 
It is interesting that two different processing methods of the same, raw FermiLAT gamma-ray data cannot be described equally well by our multi-component model. To solidify our conclusions above, further observational analysis and data collection of the GCE must be performed to determine whether the \citet{diMauro21} or \citet{Horiuchi16} data is more accurate.

\begin{figure}
	\includegraphics[width=3.5in]{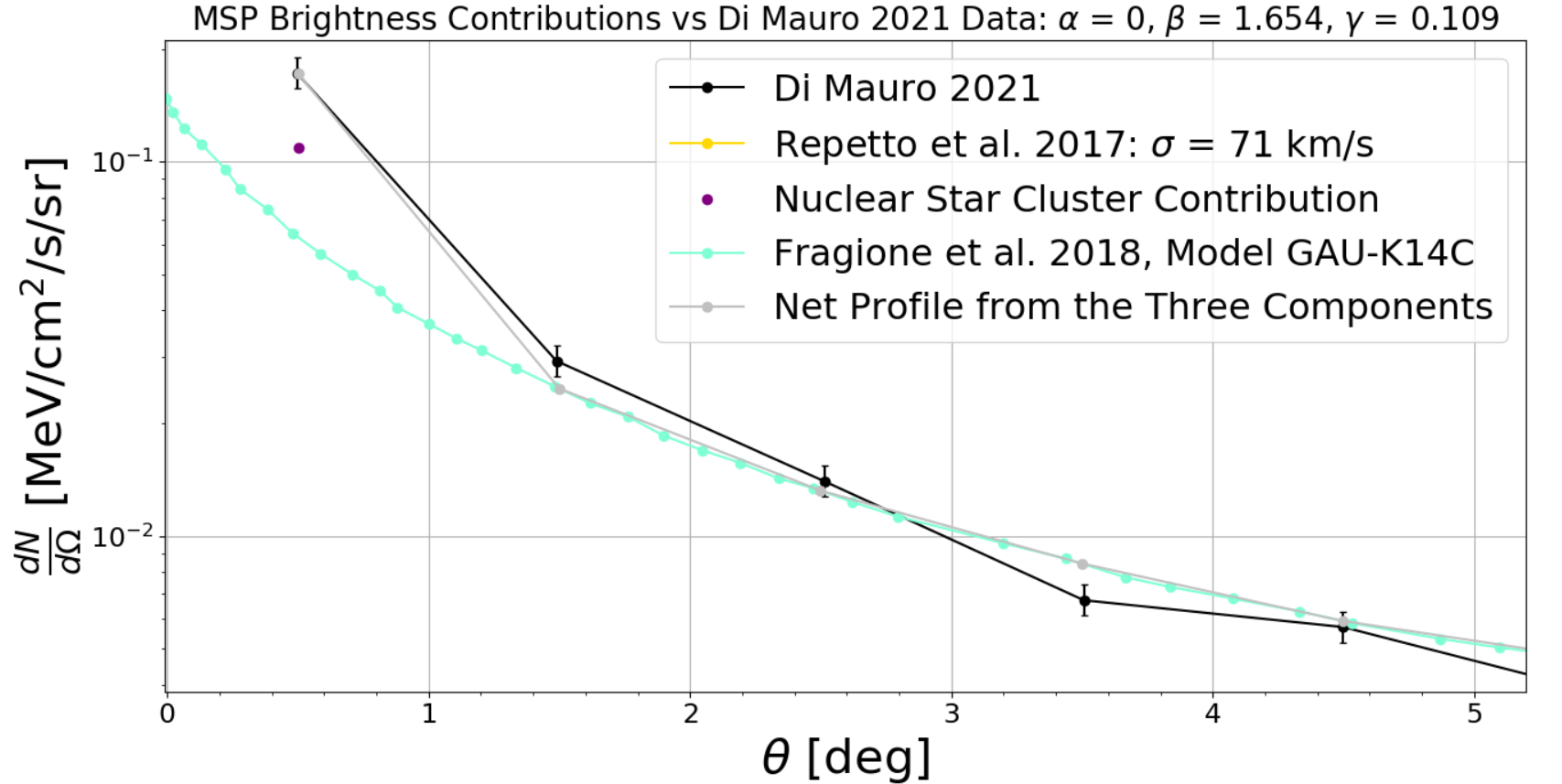}
	\includegraphics[width=3.5in]{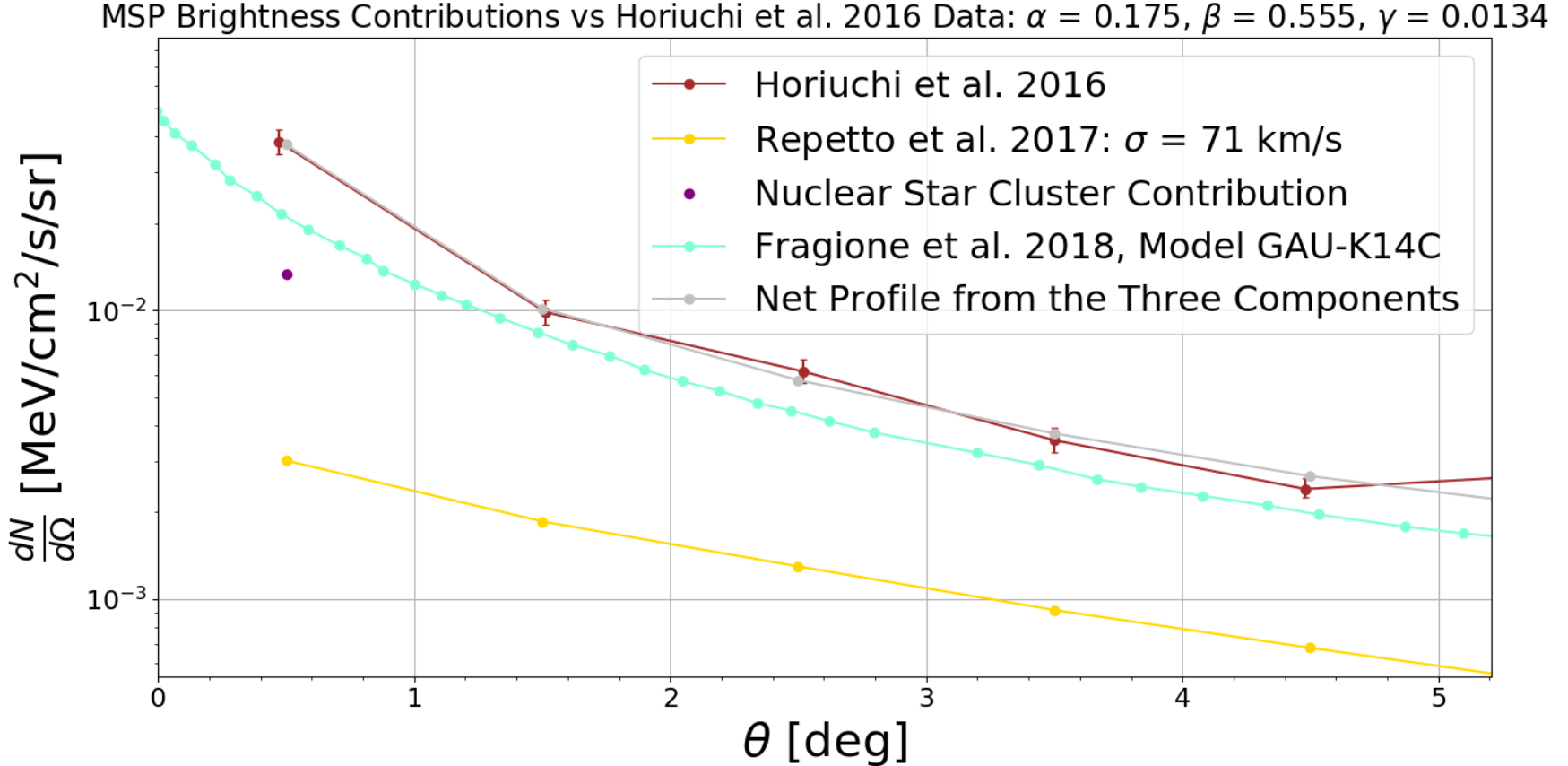}
	\includegraphics[width=3.5in]{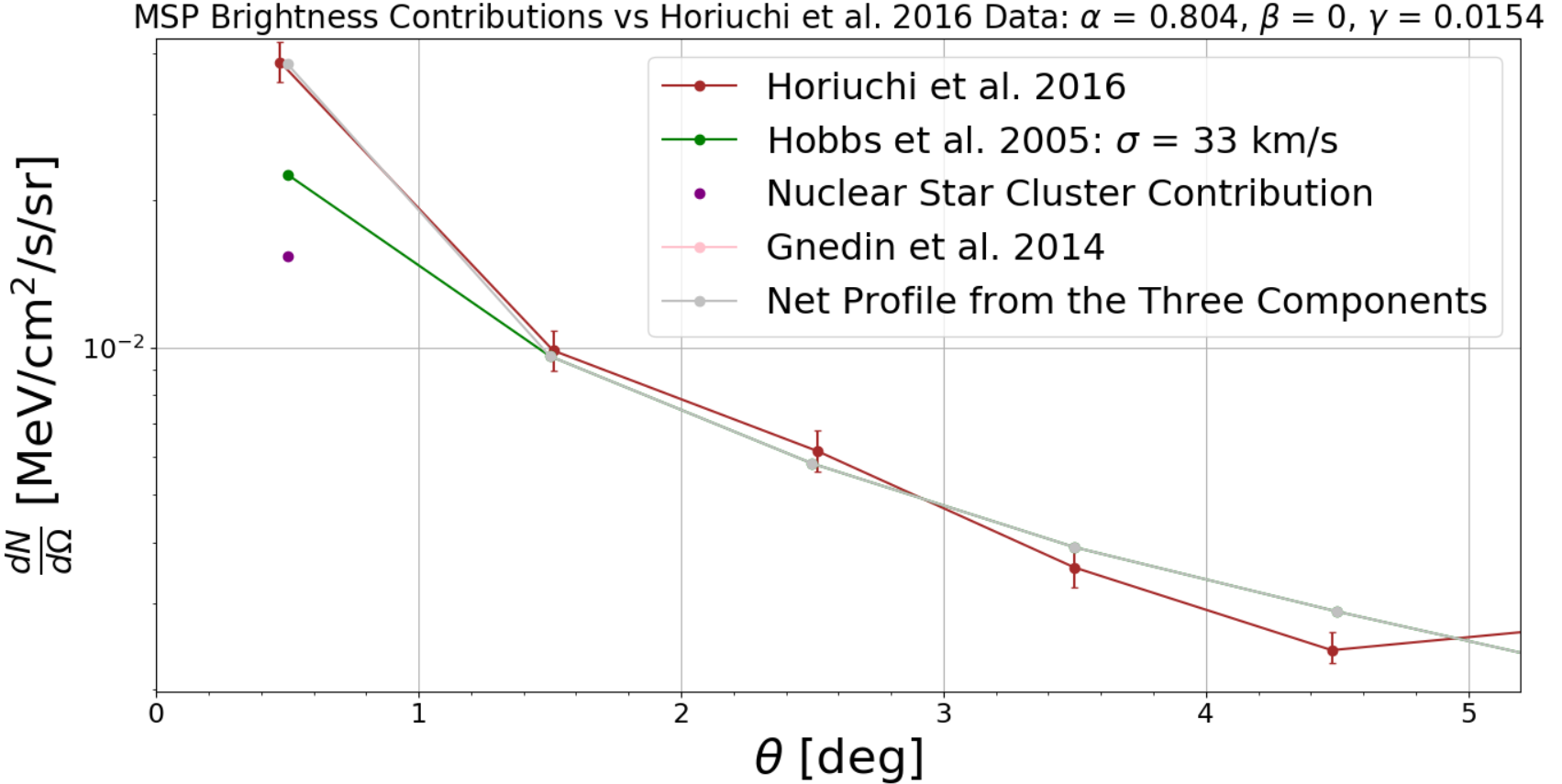}
    \caption{This figure captures three sets of best fit parameters ($\alpha$, $\beta$, $\gamma$) for varying globular cluster models and observed data sets. See text for details. 
    }
    \label{fig:msp_components}
\end{figure}

To visualize the contributions of each population, Fig.~\ref{fig:msp_components} illustrates the surface brightness profiles for each component (primordial, globular cluster deposited, and NSC) scaled by some of the best fit parameters $\alpha$, $\beta$, $\gamma$ along with the observed surface brightness profiles the best fit parameters were fit to. Further, we show an additional surface brightness profile representing the combination (summation) of the three component profiles. This illustrates the quality of the net profile resulting from the fit parameters. We 
illustrate three fits; top, an example of one of the best fits (though still quite poor, $\chi^2$/dof=8.982) to the \citet{diMauro21} data (where the primordial MSP contribution is negligible); middle, a good fit ($\chi^2$/dof=1.099) to the \citet{Horiuchi16} data (with contributions from both primordial and destroyed cluster MSPs); and bottom, a relatively poor fit ($\chi^2$/dof=1.585) to the \citet{Horiuchi16} data using primordial MSPs.
As the central-degree datapoint can be renormalized using the NSC contribution, the poor quality of the last fit is due to a mismatch at larger angles. 

\section{Conclusions}
Following observations of the Milky Way center, many analyses revealed an unexpected excess of gamma-ray emission which cannot be accounted for by astrophysical backgrounds. Two main hypotheses have surfaced to explain this gamma-ray excess. One credits the excess emission to highly concentrated annihilating dark matter particles in the Galactic Center. Another suggests an unresolved, dense population of central millisecond radio pulsars supplies the excess. 

To examine the pulsar origin as a viable explanation for the gamma-ray excess, we simulated various populations of millisecond pulsar progenitors under the influence of a realistic Milky Way galactic potential. In our N-body simulations, the pulsar populations were initially provided with various intrinsic natal kick velocities, and placed according to the known stellar mass components of the Milky Way. By tracking the pulsar trajectories for one Gyr and mapping their final positions onto the sky, we generated synthetic gamma-ray surface brightness profiles. We then 
compared the radial distributions of our simulated pulsar populations to different measurements of the gamma-ray radial profile. 
From our analysis, it appears that these Galaxy-born pulsars alone have a difficult time reproducing the concentrated brightnesses in the central 1$^\circ$. If pulsars were the source of this excess, they require incredibly low natal kick velocities to produce such concentrated emission, and still do not reproduce one of the observed gamma-ray radial profiles. 

We also investigated two alternative mechanisms for  pulsar production in the Galactic Center. Globular clusters orbiting the Milky Way can inspiral to deposit their pulsars in the central regions of the Galaxy upon their destruction. This adds to the central pulsar population leading to a peaked emission profile in the central regions of the sky. 
The other alternative is dynamical production directly within the nuclear star cluster (NSC), as seen in globular clusters.
We modeled the observed gamma-ray radial profiles with components of primordial MSPs, inspiraling evaporating globular clusters, and a NSC component. We find that the \citet{Horiuchi16} gamma-ray profile requires   a dominant component that can be either  primordial, or produced by inspiraling evaporating clusters, along with a (generally) significant (but small) NSC component. Depending on the choice of evaporating cluster model component, fits with the \citet{diMauro21} gamma-ray profile may  strongly prefer that the dominant component be produced by evaporating clusters, and also requires a significant NSC component (of order 150 MSPs). However, the \citet{diMauro21} gamma-ray fits do not achieve good fits, suggesting that if this description of the GCE is correct, then either we have not found the correct description of the MSP populations, or that (at least a large fraction of) the GCE is not produced by MSPs. 

Our analysis does not strongly prefer either a dark matter origin or a MSP origin, but it significantly constrains the MSP origin scenario. Robust determination of the gamma-ray radial profile would improve the constraints upon the possible MSP origin of the gamma-ray excess.

\section*{Acknowledgements}
We would like to thank Jo Bovy for his useful galactic dynamics package \textit{galpy} and insightful discussions regarding inspiraling globular clusters, and J. Bramante for discussions on related topics.
COH is supported by NSERC Discovery Grant RGPIN-2016-04602, and OB in part by an NSERC USRA.

\section*{Data Availability}
The data underlying this article will be shared on reasonable request to the corresponding author.


\bibliographystyle{mnras}
\bibliography{boodram} 

\begin{thebibliography}{}
\makeatletter
\relax
\def\mn@urlcharsother{\let\do\@makeother \do\$\do\&\do\#\do\^\do\_\do\%\do\~}
\def\mn@doi{\begingroup\mn@urlcharsother \@ifnextchar [ {\mn@doi@}
  {\mn@doi@[]}}
\def\mn@doi@[#1]#2{\def\@tempa{#1}\ifx\@tempa\@empty \href
  {http://dx.doi.org/#2} {doi:#2}\else \href {http://dx.doi.org/#2} {#1}\fi
  \endgroup}
\def\mn@eprint#1#2{\mn@eprint@#1:#2::\@nil}
\def\mn@eprint@arXiv#1{\href {http://arxiv.org/abs/#1} {{\tt arXiv:#1}}}
\def\mn@eprint@dblp#1{\href {http://dblp.uni-trier.de/rec/bibtex/#1.xml}
  {dblp:#1}}
\def\mn@eprint@#1:#2:#3:#4\@nil{\def\@tempa {#1}\def\@tempb {#2}\def\@tempc
  {#3}\ifx \@tempc \@empty \let \@tempc \@tempb \let \@tempb \@tempa \fi \ifx
  \@tempb \@empty \def\@tempb {arXiv}\fi \@ifundefined
  {mn@eprint@\@tempb}{\@tempb:\@tempc}{\expandafter \expandafter \csname
  mn@eprint@\@tempb\endcsname \expandafter{\@tempc}}}

\bibitem[\protect\citeauthoryear{{Abazajian}}{{Abazajian}}{2011}]{Abazajian11}
{Abazajian} K.~N.,  2011, \mn@doi [\jcap] {10.1088/1475-7516/2011/03/010},
  \href {https://ui.adsabs.harvard.edu/abs/2011JCAP...03..010A} {2011, 010}

\bibitem[\protect\citeauthoryear{{Abazajian} \& {Kaplinghat}}{{Abazajian} \&
  {Kaplinghat}}{2012}]{Abazajian12}
{Abazajian} K.~N.,  {Kaplinghat} M.,  2012, \mn@doi [\prd]
  {10.1103/PhysRevD.86.083511}, \href
  {https://ui.adsabs.harvard.edu/abs/2012PhRvD..86h3511A} {86, 083511}

\bibitem[\protect\citeauthoryear{{Abbate}, {Mastrobuono-Battisti}, {Colpi},
  {Possenti}, {Sippel}  \& {Dotti}}{{Abbate} et~al.}{2018}]{Abbate18}
{Abbate} F.,  {Mastrobuono-Battisti} A.,  {Colpi} M.,  {Possenti} A.,  {Sippel}
  A.~C.,   {Dotti} M.,  2018, \mn@doi [\mnras] {10.1093/mnras/stx2364}, \href
  {https://ui.adsabs.harvard.edu/abs/2018MNRAS.473..927A} {473, 927}

\bibitem[\protect\citeauthoryear{{Abdo} et~al.,}{{Abdo} et~al.}{2009}]{Abdo09}
{Abdo} A.~A.,  et~al., 2009, \mn@doi [Science] {10.1126/science.1176113}, \href
  {https://ui.adsabs.harvard.edu/abs/2009Sci...325..848A} {325, 848}

\bibitem[\protect\citeauthoryear{{Abdo} et~al.,}{{Abdo} et~al.}{2010}]{Abdo10}
{Abdo} A.~A.,  et~al., 2010, \mn@doi [\aap] {10.1051/0004-6361/201014458},
  \href {https://ui.adsabs.harvard.edu/abs/2010A&A...524A..75A} {524, A75}

\bibitem[\protect\citeauthoryear{{Ackermann} et~al.,}{{Ackermann}
  et~al.}{2017}]{Ackermann17}
{Ackermann} M.,  et~al., 2017, \mn@doi [\apj] {10.3847/1538-4357/aa6cab}, \href
  {https://ui.adsabs.harvard.edu/abs/2017ApJ...840...43A} {840, 43}

\bibitem[\protect\citeauthoryear{{Ajello} et~al.,}{{Ajello}
  et~al.}{2016}]{Ajello16}
{Ajello} M.,  et~al., 2016, \mn@doi [\apj] {10.3847/0004-637X/819/1/44}, \href
  {https://ui.adsabs.harvard.edu/abs/2016ApJ...819...44A} {819, 44}

\bibitem[\protect\citeauthoryear{{Alpar}, {Cheng}, {Ruderman}  \&
  {Shaham}}{{Alpar} et~al.}{1982}]{Alpar82}
{Alpar} M.~A.,  {Cheng} A.~F.,  {Ruderman} M.~A.,   {Shaham} J.,  1982, \mn@doi
  [\nat] {10.1038/300728a0}, \href
  {https://ui.adsabs.harvard.edu/abs/1982Natur.300..728A} {300, 728}

\bibitem[\protect\citeauthoryear{{Archibald} et~al.,}{{Archibald}
  et~al.}{2009}]{Archibald09}
{Archibald} A.~M.,  et~al., 2009, \mn@doi [Science] {10.1126/science.1172740},
  \href {https://ui.adsabs.harvard.edu/abs/2009Sci...324.1411A} {324, 1411}

\bibitem[\protect\citeauthoryear{{Astropy Collaboration} et~al.,}{{Astropy
  Collaboration} et~al.}{2018}]{Astropy2018}
{Astropy Collaboration} et~al., 2018, \mn@doi [\aj] {10.3847/1538-3881/aabc4f},
  \href {https://ui.adsabs.harvard.edu/abs/2018AJ....156..123A} {156, 123}

\bibitem[\protect\citeauthoryear{{Bagchi}, {Lorimer}  \&
  {Chennamangalam}}{{Bagchi} et~al.}{2011}]{Bagchi11}
{Bagchi} M.,  {Lorimer} D.~R.,   {Chennamangalam} J.,  2011, \mn@doi [\mnras]
  {10.1111/j.1365-2966.2011.19498.x}, \href
  {https://ui.adsabs.harvard.edu/abs/2011MNRAS.418..477B} {418, 477}

\bibitem[\protect\citeauthoryear{{Bahramian}, {Heinke}, {Sivakoff}  \&
  {Gladstone}}{{Bahramian} et~al.}{2013}]{Bahramian13}
{Bahramian} A.,  {Heinke} C.~O.,  {Sivakoff} G.~R.,   {Gladstone} J.~C.,  2013,
  \mn@doi [\apj] {10.1088/0004-637X/766/2/136}, \href
  {https://ui.adsabs.harvard.edu/abs/2013ApJ...766..136B} {766, 136}

\bibitem[\protect\citeauthoryear{{Bartels}, {Storm}, {Weniger}  \&
  {Calore}}{{Bartels} et~al.}{2018}]{Bartels18}
{Bartels} R.,  {Storm} E.,  {Weniger} C.,   {Calore} F.,  2018, \mn@doi [Nature
  Astronomy] {10.1038/s41550-018-0531-z}, \href
  {https://ui.adsabs.harvard.edu/abs/2018NatAs...2..819B} {2, 819}

\bibitem[\protect\citeauthoryear{{Bhattacharya} \& {van den
  Heuvel}}{{Bhattacharya} \& {van den Heuvel}}{1991}]{Bhattacharya91}
{Bhattacharya} D.,  {van den Heuvel} E.~P.~J.,  1991, \mn@doi [\physrep]
  {10.1016/0370-1573(91)90064-S}, \href
  {https://ui.adsabs.harvard.edu/abs/1991PhR...203....1B} {203, 1}

\bibitem[\protect\citeauthoryear{{Bland-Hawthorn} \&
  {Gerhard}}{{Bland-Hawthorn} \& {Gerhard}}{2016}]{Bland-Hawthorn16}
{Bland-Hawthorn} J.,  {Gerhard} O.,  2016, \mn@doi [\araa]
  {10.1146/annurev-astro-081915-023441}, \href
  {https://ui.adsabs.harvard.edu/abs/2016ARA&A..54..529B} {54, 529}

\bibitem[\protect\citeauthoryear{{Bovy}}{{Bovy}}{2015}]{Bovy15}
{Bovy} J.,  2015, \mn@doi [\apjs] {10.1088/0067-0049/216/2/29}, \href
  {https://ui.adsabs.harvard.edu/abs/2015ApJS..216...29B} {216, 29}

\bibitem[\protect\citeauthoryear{{Brandt} \& {Kocsis}}{{Brandt} \&
  {Kocsis}}{2015}]{Brandt15}
{Brandt} T.~D.,  {Kocsis} B.,  2015, \mn@doi [\apj]
  {10.1088/0004-637X/812/1/15}, \href
  {https://ui.adsabs.harvard.edu/abs/2015ApJ...812...15B} {812, 15}

\bibitem[\protect\citeauthoryear{{Brandt} \& {Podsiadlowski}}{{Brandt} \&
  {Podsiadlowski}}{1995}]{Brandt95}
{Brandt} N.,  {Podsiadlowski} P.,  1995, \mn@doi [\mnras]
  {10.1093/mnras/274.2.461}, \href
  {https://ui.adsabs.harvard.edu/abs/1995MNRAS.274..461B} {274, 461}

\bibitem[\protect\citeauthoryear{{Buschmann}, {Rodd}, {Safdi}, {Chang},
  {Mishra-Sharma}, {Lisanti}  \& {Macias}}{{Buschmann}
  et~al.}{2020}]{Buschmann20}
{Buschmann} M.,  {Rodd} N.~L.,  {Safdi} B.~R.,  {Chang} L.~J.,  {Mishra-Sharma}
  S.,  {Lisanti} M.,   {Macias} O.,  2020, \mn@doi [\prd]
  {10.1103/PhysRevD.102.023023}, \href
  {https://ui.adsabs.harvard.edu/abs/2020PhRvD.102b3023B} {102, 023023}

\bibitem[\protect\citeauthoryear{{Camilo} \& {Rasio}}{{Camilo} \&
  {Rasio}}{2005}]{Camilo05}
{Camilo} F.,  {Rasio} F.~A.,  2005, in {Rasio} F.~A.,  {Stairs} I.~H.,  eds,
  Astronomical Society of the Pacific Conference Series Vol. 328, Binary Radio
  Pulsars. p.~147 (\mn@eprint {arXiv} {astro-ph/0501226})

\bibitem[\protect\citeauthoryear{{Carlson} \& {Profumo}}{{Carlson} \&
  {Profumo}}{2014}]{Carlson14}
{Carlson} E.,  {Profumo} S.,  2014, \mn@doi [\prd]
  {10.1103/PhysRevD.90.023015}, \href
  {https://ui.adsabs.harvard.edu/abs/2014PhRvD..90b3015C} {90, 023015}

\bibitem[\protect\citeauthoryear{{Cerde{\~n}o}, {Peir{\'o}}  \&
  {Robles}}{{Cerde{\~n}o} et~al.}{2015}]{Cerdeno15}
{Cerde{\~n}o} D.~G.,  {Peir{\'o}} M.,   {Robles} S.,  2015, \mn@doi [\prd]
  {10.1103/PhysRevD.91.123530}, \href
  {https://ui.adsabs.harvard.edu/abs/2015PhRvD..91l3530C} {91, 123530}

\bibitem[\protect\citeauthoryear{{Cholis}, {Hooper}  \& {Linden}}{{Cholis}
  et~al.}{2015a}]{Cholis15}
{Cholis} I.,  {Hooper} D.,   {Linden} T.,  2015a, \mn@doi [\jcap]
  {10.1088/1475-7516/2015/06/043}, \href
  {https://ui.adsabs.harvard.edu/abs/2015JCAP...06..043C} {2015, 043}

\bibitem[\protect\citeauthoryear{{Cholis}, {Evoli}, {Calore}, {Linden},
  {Weniger}  \& {Hooper}}{{Cholis} et~al.}{2015b}]{Cholis15b}
{Cholis} I.,  {Evoli} C.,  {Calore} F.,  {Linden} T.,  {Weniger} C.,   {Hooper}
  D.,  2015b, \mn@doi [\jcap] {10.1088/1475-7516/2015/12/005}, \href
  {https://ui.adsabs.harvard.edu/abs/2015JCAP...12..005C} {2015, 005}

\bibitem[\protect\citeauthoryear{{Daylan}, {Finkbeiner}, {Hooper}, {Linden},
  {Portillo}, {Rodd}  \& {Slatyer}}{{Daylan} et~al.}{2016}]{Daylan16}
{Daylan} T.,  {Finkbeiner} D.~P.,  {Hooper} D.,  {Linden} T.,  {Portillo} S.
  K.~N.,  {Rodd} N.~L.,   {Slatyer} T.~R.,  2016, \mn@doi [Physics of the Dark
  Universe] {10.1016/j.dark.2015.12.005}, \href
  {https://ui.adsabs.harvard.edu/abs/2016PDU....12....1D} {12, 1}

\bibitem[\protect\citeauthoryear{{Di Mauro}}{{Di Mauro}}{2021}]{diMauro21}
{Di Mauro} M.,  2021, \mn@doi [\prd] {10.1103/PhysRevD.103.063029}, \href
  {https://ui.adsabs.harvard.edu/abs/2021PhRvD.103f3029D} {103, 063029}

\bibitem[\protect\citeauthoryear{{Eckner} et~al.,}{{Eckner}
  et~al.}{2018}]{Eckner18}
{Eckner} C.,  et~al., 2018, \mn@doi [\apj] {10.3847/1538-4357/aac029}, \href
  {https://ui.adsabs.harvard.edu/abs/2018ApJ...862...79E} {862, 79}

\bibitem[\protect\citeauthoryear{{Faucher-Gigu{\`e}re} \&
  {Loeb}}{{Faucher-Gigu{\`e}re} \& {Loeb}}{2011}]{Faucher-Giguere11}
{Faucher-Gigu{\`e}re} C.-A.,  {Loeb} A.,  2011, \mn@doi [\mnras]
  {10.1111/j.1365-2966.2011.19019.x}, \href
  {https://ui.adsabs.harvard.edu/abs/2011MNRAS.415.3951F} {415, 3951}

\bibitem[\protect\citeauthoryear{{Fragione}, {Antonini}  \&
  {Gnedin}}{{Fragione} et~al.}{2018}]{Fragione18}
{Fragione} G.,  {Antonini} F.,   {Gnedin} O.~Y.,  2018, \mn@doi [\mnras]
  {10.1093/mnras/sty183}, \href
  {https://ui.adsabs.harvard.edu/abs/2018MNRAS.475.5313F} {475, 5313}

\bibitem[\protect\citeauthoryear{{Generozov}, {Stone}, {Metzger}  \&
  {Ostriker}}{{Generozov} et~al.}{2018}]{Generozov18}
{Generozov} A.,  {Stone} N.~C.,  {Metzger} B.~D.,   {Ostriker} J.~P.,  2018,
  \mn@doi [\mnras] {10.1093/mnras/sty1262}, \href
  {https://ui.adsabs.harvard.edu/abs/2018MNRAS.478.4030G} {478, 4030}

\bibitem[\protect\citeauthoryear{{Gnedin}, {Ostriker}  \& {Tremaine}}{{Gnedin}
  et~al.}{2014}]{Gnedin14}
{Gnedin} O.~Y.,  {Ostriker} J.~P.,   {Tremaine} S.,  2014, \mn@doi [\apj]
  {10.1088/0004-637X/785/1/71}, \href
  {https://ui.adsabs.harvard.edu/abs/2014ApJ...785...71G} {785, 71}

\bibitem[\protect\citeauthoryear{{Gonthier}, {Harding}, {Ferrara}, {Frederick},
  {Mohr}  \& {Koh}}{{Gonthier} et~al.}{2018}]{Gonthier18}
{Gonthier} P.~L.,  {Harding} A.~K.,  {Ferrara} E.~C.,  {Frederick} S.~E.,
  {Mohr} V.~E.,   {Koh} Y.-M.,  2018, \mn@doi [\apj]
  {10.3847/1538-4357/aad08d}, \href
  {https://ui.adsabs.harvard.edu/abs/2018ApJ...863..199G} {863, 199}

\bibitem[\protect\citeauthoryear{{Goodenough} \& {Hooper}}{{Goodenough} \&
  {Hooper}}{2009}]{Goodenough09}
{Goodenough} L.,  {Hooper} D.,  2009, arXiv e-prints, \href
  {https://ui.adsabs.harvard.edu/abs/2009arXiv0910.2998G} {p. arXiv:0910.2998}

\bibitem[\protect\citeauthoryear{{Heinke}, {Grindlay}, {Edmonds}, {Cohn},
  {Lugger}, {Camilo}, {Bogdanov}  \& {Freire}}{{Heinke}
  et~al.}{2005}]{Heinke05}
{Heinke} C.~O.,  {Grindlay} J.~E.,  {Edmonds} P.~D.,  {Cohn} H.~N.,  {Lugger}
  P.~M.,  {Camilo} F.,  {Bogdanov} S.,   {Freire} P.~C.,  2005, \mn@doi [\apj]
  {10.1086/429899}, \href
  {https://ui.adsabs.harvard.edu/abs/2005ApJ...625..796H} {625, 796}

\bibitem[\protect\citeauthoryear{{Hobbs}, {Lorimer}, {Lyne}  \&
  {Kramer}}{{Hobbs} et~al.}{2005}]{Hobbs05}
{Hobbs} G.,  {Lorimer} D.~R.,  {Lyne} A.~G.,   {Kramer} M.,  2005, \mn@doi
  [\mnras] {10.1111/j.1365-2966.2005.09087.x}, \href
  {https://ui.adsabs.harvard.edu/abs/2005MNRAS.360..974H} {360, 974}

\bibitem[\protect\citeauthoryear{{Hooper} \& {Goodenough}}{{Hooper} \&
  {Goodenough}}{2011}]{Hooper11}
{Hooper} D.,  {Goodenough} L.,  2011, \mn@doi [Physics Letters B]
  {10.1016/j.physletb.2011.02.029}, \href
  {https://ui.adsabs.harvard.edu/abs/2011PhLB..697..412H} {697, 412}

\bibitem[\protect\citeauthoryear{{Hooper} \& {Linden}}{{Hooper} \&
  {Linden}}{2011}]{Hooper11b}
{Hooper} D.,  {Linden} T.,  2011, \mn@doi [\prd] {10.1103/PhysRevD.84.123005},
  \href {https://ui.adsabs.harvard.edu/abs/2011PhRvD..84l3005H} {84, 123005}

\bibitem[\protect\citeauthoryear{{Hooper} \& {Mohlabeng}}{{Hooper} \&
  {Mohlabeng}}{2016}]{Hooper16}
{Hooper} D.,  {Mohlabeng} G.,  2016, \mn@doi [\jcap]
  {10.1088/1475-7516/2016/03/049}, \href
  {https://ui.adsabs.harvard.edu/abs/2016JCAP...03..049H} {2016, 049}

\bibitem[\protect\citeauthoryear{{Horiuchi}, {Kaplinghat}  \& {Kwa}}{{Horiuchi}
  et~al.}{2016}]{Horiuchi16}
{Horiuchi} S.,  {Kaplinghat} M.,   {Kwa} A.,  2016, \mn@doi [\jcap]
  {10.1088/1475-7516/2016/11/053}, \href
  {https://ui.adsabs.harvard.edu/abs/2016JCAP...11..053H} {2016, 053}

\bibitem[\protect\citeauthoryear{{Horta} et~al.,}{{Horta}
  et~al.}{2021}]{Horta21}
{Horta} D.,  et~al., 2021, \mn@doi [\mnras] {10.1093/mnras/staa3598}, \href
  {https://ui.adsabs.harvard.edu/abs/2021MNRAS.500.5462H} {500, 5462}

\bibitem[\protect\citeauthoryear{{Hui}, {Cheng}, {Wang}, {Tam}, {Kong},
  {Chernyshov}  \& {Dogiel}}{{Hui} et~al.}{2011}]{Hui11}
{Hui} C.~Y.,  {Cheng} K.~S.,  {Wang} Y.,  {Tam} P.~H.~T.,  {Kong} A.~K.~H.,
  {Chernyshov} D.~O.,   {Dogiel} V.~A.,  2011, \mn@doi [\apj]
  {10.1088/0004-637X/726/2/100}, \href
  {https://ui.adsabs.harvard.edu/abs/2011ApJ...726..100H} {726, 100}

\bibitem[\protect\citeauthoryear{{Jonker} \& {Nelemans}}{{Jonker} \&
  {Nelemans}}{2004}]{Jonker04}
{Jonker} P.~G.,  {Nelemans} G.,  2004, \mn@doi [\mnras]
  {10.1111/j.1365-2966.2004.08193.x}, \href
  {https://ui.adsabs.harvard.edu/abs/2004MNRAS.354..355J} {354, 355}

\bibitem[\protect\citeauthoryear{{Launhardt}, {Zylka}  \& {Mezger}}{{Launhardt}
  et~al.}{2002}]{Launhardt02}
{Launhardt} R.,  {Zylka} R.,   {Mezger} P.~G.,  2002, \mn@doi [\aap]
  {10.1051/0004-6361:20020017}, \href
  {https://ui.adsabs.harvard.edu/abs/2002A&A...384..112L} {384, 112}

\bibitem[\protect\citeauthoryear{{Leane} \& {Slatyer}}{{Leane} \&
  {Slatyer}}{2019}]{Leane19}
{Leane} R.~K.,  {Slatyer} T.~R.,  2019, \mn@doi [\prl]
  {10.1103/PhysRevLett.123.241101}, \href
  {https://ui.adsabs.harvard.edu/abs/2019PhRvL.123x1101L} {123, 241101}

\bibitem[\protect\citeauthoryear{{Leane} \& {Slatyer}}{{Leane} \&
  {Slatyer}}{2020}]{Leane20}
{Leane} R.~K.,  {Slatyer} T.~R.,  2020, \mn@doi [\prl]
  {10.1103/PhysRevLett.125.121105}, \href
  {https://ui.adsabs.harvard.edu/abs/2020PhRvL.125l1105L} {125, 121105}

\bibitem[\protect\citeauthoryear{{Lee}, {Lisanti}, {Safdi}, {Slatyer}  \&
  {Xue}}{{Lee} et~al.}{2016}]{Lee16}
{Lee} S.~K.,  {Lisanti} M.,  {Safdi} B.~R.,  {Slatyer} T.~R.,   {Xue} W.,
  2016, \mn@doi [\prl] {10.1103/PhysRevLett.116.051103}, \href
  {https://ui.adsabs.harvard.edu/abs/2016PhRvL.116e1103L} {116, 051103}

\bibitem[\protect\citeauthoryear{{Macias} \& {Gordon}}{{Macias} \&
  {Gordon}}{2014}]{Macias14}
{Macias} O.,  {Gordon} C.,  2014, \mn@doi [\prd] {10.1103/PhysRevD.89.063515},
  \href {https://ui.adsabs.harvard.edu/abs/2014PhRvD..89f3515M} {89, 063515}

\bibitem[\protect\citeauthoryear{{Macias}, {Gordon}, {Crocker}, {Coleman},
  {Paterson}, {Horiuchi}  \& {Pohl}}{{Macias} et~al.}{2018}]{Macias18}
{Macias} O.,  {Gordon} C.,  {Crocker} R.~M.,  {Coleman} B.,  {Paterson} D.,
  {Horiuchi} S.,   {Pohl} M.,  2018, \mn@doi [Nature Astronomy]
  {10.1038/s41550-018-0414-3}, \href
  {https://ui.adsabs.harvard.edu/abs/2018NatAs...2..387M} {2, 387}

\bibitem[\protect\citeauthoryear{{Macias}, {Horiuchi}, {Kaplinghat}, {Gordon},
  {Crocker}  \& {Nataf}}{{Macias} et~al.}{2019}]{Macias19}
{Macias} O.,  {Horiuchi} S.,  {Kaplinghat} M.,  {Gordon} C.,  {Crocker} R.~M.,
   {Nataf} D.~M.,  2019, \mn@doi [\jcap] {10.1088/1475-7516/2019/09/042}, \href
  {https://ui.adsabs.harvard.edu/abs/2019JCAP...09..042M} {2019, 042}

\bibitem[\protect\citeauthoryear{{Muno}, {Pfahl}, {Baganoff}, {Brandt}, {Ghez},
  {Lu}  \& {Morris}}{{Muno} et~al.}{2005}]{Muno05}
{Muno} M.~P.,  {Pfahl} E.,  {Baganoff} F.~K.,  {Brandt} W.~N.,  {Ghez} A.,
  {Lu} J.,   {Morris} M.~R.,  2005, \mn@doi [\apjl] {10.1086/429721}, \href
  {https://ui.adsabs.harvard.edu/abs/2005ApJ...622L.113M} {622, L113}

\bibitem[\protect\citeauthoryear{{Papitto} et~al.,}{{Papitto}
  et~al.}{2013}]{Papitto13}
{Papitto} A.,  et~al., 2013, \mn@doi [\nat] {10.1038/nature12470}, \href
  {https://ui.adsabs.harvard.edu/abs/2013Natur.501..517P} {501, 517}

\bibitem[\protect\citeauthoryear{{Pflamm-Altenburg} \&
  {Kroupa}}{{Pflamm-Altenburg} \& {Kroupa}}{2009}]{Pflamm-Altenburg09}
{Pflamm-Altenburg} J.,  {Kroupa} P.,  2009, \mn@doi [\mnras]
  {10.1111/j.1365-2966.2009.14954.x}, \href
  {https://ui.adsabs.harvard.edu/abs/2009MNRAS.397..488P} {397, 488}

\bibitem[\protect\citeauthoryear{{Phinney} \& {Kulkarni}}{{Phinney} \&
  {Kulkarni}}{1994}]{Phinney94}
{Phinney} E.~S.,  {Kulkarni} S.~R.,  1994, \mn@doi [\araa]
  {10.1146/annurev.aa.32.090194.003111}, \href
  {https://ui.adsabs.harvard.edu/abs/1994ARA&A..32..591P} {32, 591}

\bibitem[\protect\citeauthoryear{{Ploeg} \& {Gordon}}{{Ploeg} \&
  {Gordon}}{2021}]{Ploeg21}
{Ploeg} H.,  {Gordon} C.,  2021, arXiv e-prints, \href
  {https://ui.adsabs.harvard.edu/abs/2021arXiv210513034P} {p. arXiv:2105.13034}

\bibitem[\protect\citeauthoryear{{Ploeg}, {Gordon}, {Crocker}  \&
  {Macias}}{{Ploeg} et~al.}{2017}]{Ploeg17}
{Ploeg} H.,  {Gordon} C.,  {Crocker} R.,   {Macias} O.,  2017, \mn@doi [\jcap]
  {10.1088/1475-7516/2017/08/015}, \href
  {https://ui.adsabs.harvard.edu/abs/2017JCAP...08..015P} {2017, 015}

\bibitem[\protect\citeauthoryear{{Repetto}, {Igoshev}  \& {Nelemans}}{{Repetto}
  et~al.}{2017}]{Repetto17}
{Repetto} S.,  {Igoshev} A.~P.,   {Nelemans} G.,  2017, \mn@doi [\mnras]
  {10.1093/mnras/stx027}, \href
  {https://ui.adsabs.harvard.edu/abs/2017MNRAS.467..298R} {467, 298}

\bibitem[\protect\citeauthoryear{{Sch{\"o}del}, {Merritt}  \&
  {Eckart}}{{Sch{\"o}del} et~al.}{2009}]{Schodel09}
{Sch{\"o}del} R.,  {Merritt} D.,   {Eckart} A.,  2009, \mn@doi [\aap]
  {10.1051/0004-6361/200810922}, \href
  {https://ui.adsabs.harvard.edu/abs/2009A&A...502...91S} {502, 91}

\bibitem[\protect\citeauthoryear{{Turk} \& {Lorimer}}{{Turk} \&
  {Lorimer}}{2013}]{Turk13}
{Turk} P.~J.,  {Lorimer} D.~R.,  2013, \mn@doi [\mnras]
  {10.1093/mnras/stt1850}, \href
  {https://ui.adsabs.harvard.edu/abs/2013MNRAS.436.3720T} {436, 3720}

\bibitem[\protect\citeauthoryear{{Valenti} et~al.,}{{Valenti}
  et~al.}{2018}]{Valenti18}
{Valenti} E.,  et~al., 2018, \mn@doi [\aap] {10.1051/0004-6361/201832905},
  \href {https://ui.adsabs.harvard.edu/abs/2018A&A...616A..83V} {616, A83}

\bibitem[\protect\citeauthoryear{{Vitale} \& {Morselli}}{{Vitale} \&
  {Morselli}}{2009}]{Vitale09}
{Vitale} V.,  {Morselli} A.,  2009, arXiv e-prints, \href
  {https://ui.adsabs.harvard.edu/abs/2009arXiv0912.3828V} {p. arXiv:0912.3828}

\bibitem[\protect\citeauthoryear{{Voss} \& {Gilfanov}}{{Voss} \&
  {Gilfanov}}{2007a}]{Voss07b}
{Voss} R.,  {Gilfanov} M.,  2007a, \mn@doi [\mnras]
  {10.1111/j.1365-2966.2007.12223.x}, \href
  {https://ui.adsabs.harvard.edu/abs/2007MNRAS.380.1685V} {380, 1685}

\bibitem[\protect\citeauthoryear{{Voss} \& {Gilfanov}}{{Voss} \&
  {Gilfanov}}{2007b}]{Voss07a}
{Voss} R.,  {Gilfanov} M.,  2007b, \mn@doi [\aap] {10.1051/0004-6361:20066614},
  \href {https://ui.adsabs.harvard.edu/abs/2007A&A...468...49V} {468, 49}

\bibitem[\protect\citeauthoryear{{Xue}, {Rix}, {Ma}, {Morrison}, {Bovy},
  {Sesar}  \& {Janesh}}{{Xue} et~al.}{2015}]{Xue15}
{Xue} X.-X.,  {Rix} H.-W.,  {Ma} Z.,  {Morrison} H.,  {Bovy} J.,  {Sesar} B.,
  {Janesh} W.,  2015, \mn@doi [\apj] {10.1088/0004-637X/809/2/144}, \href
  {https://ui.adsabs.harvard.edu/abs/2015ApJ...809..144X} {809, 144}

\bibitem[\protect\citeauthoryear{{Yuan} \& {Zhang}}{{Yuan} \&
  {Zhang}}{2014}]{Yuan14}
{Yuan} Q.,  {Zhang} B.,  2014, \mn@doi [Journal of High Energy Astrophysics]
  {10.1016/j.jheap.2014.06.001}, \href
  {https://ui.adsabs.harvard.edu/abs/2014JHEAp...3....1Y} {3, 1}

\makeatother
\end{thebibliography}

\bsp	
\label{lastpage}
\end{document}